\documentclass[
  journal=largetwo,
  manuscript=article-type,
  year=2023,
  volume=37,
]{cup-journal}

\usepackage{amsmath}
\usepackage[nopatch]{microtype}
\usepackage{booktabs}

\usepackage{graphicx}	
\usepackage{amsmath}	
\usepackage{amssymb}	

\usepackage{mathptmx}
\usepackage{txfonts}
\usepackage[T1]{fontenc}
\usepackage{hyperref}

\title{HINORA, a method for detecting ring-like structures in 3D point distributions I: application to the Local Volume Galaxy catalogue}

\author{Edward Olex}
\affiliation{Departamento de F\'isica Te\'{o}rica, M\'{o}dulo 15, Facultad de Ciencias, Universidad Aut\'{o}noma de Madrid, 28049 Madrid, Spain}
\email[E. Olex]{edward.olex@estudiante.uam.es}

\author{Alexander Knebe}
\affiliation{Departamento de F\'isica Te\'{o}rica, M\'{o}dulo 15, Facultad de Ciencias, Universidad Aut\'{o}noma de Madrid, 28049 Madrid, Spain}
\alsoaffiliation{Centro de Investigaci\'{o}n Avanzada en F\'isica Fundamental (CIAFF), Facultad de Ciencias, Universidad Aut\'{o}noma de Madrid, 28049 Madrid, Spain}
\alsoaffiliation{International Centre for Radio Astronomy Research, University of Western Australia, 35 Stirling Highway, Crawley, Western Australia 6009, Australia}

\author{Noam I. Libeskind}
\affiliation{Leibniz-Institut f\"ur Astrophysik Potsdam (AIP), An der Sternwarte 16, D-14482 Potsdam, Germany}

\author{Dmitry I. Makarov}
\affiliation{Special Astrophysical Observatory, Russian Academy of Sciences, Nizhnii Arkhyz, 369167 Russia}

\author{Stefan Gottl\"ober}
\affiliation{Leibniz-Institut f\"ur Astrophysik Potsdam (AIP), An der Sternwarte 16, D-14482 Potsdam, Germany}


\begin{document}

\begin{abstract}
We present a new method -- called HINORA (HIgh-NOise RANdom SAmple Consensus) -- for the identification of regular structures in 3D point distributions. Motivated by the possible existence of the so-called Council of Giants, i.e. a ring of twelve massive galaxies surrounding the Local Group in the Local Sheet with a radius of 3.75 Mpc, we apply HINORA to the Local Volume Galaxy catalogue confirming its existence. When varying the lower limit of K-band luminosity of the galaxy entering the catalogue, we further report on the existence of another ring-like structure in the Local Volume that now contains the Milky Way and M31. However, this newly found structure is dominated by low-mass (satellite) galaxies. While we here simply present the novel method as well as its first application to observational data, follow-up work using numerical simulations of cosmic structure formation shall shed light into the origin of such regular patterns in the galaxy distribution. Further, the method is equally suited to identify similar (or even different) structures in various kinds of astrophysical data (e.g. locating the actual `baryonic-acoustic oscillation spheres' in galaxy redshift surveys).
\end{abstract}

\section{Introduction}
Understanding the process of cosmic structure formation enables us to unravel the origins and evolution of the Universe and provides valuable insights into the fundamental laws that govern its behavior. The first surveys that accurately cataloged the sky by measuring distances \citep[e.g.][]{Huchra1983,Geller1989} noted a non-uniform distribution, cosmic web-like structure. Subsequent surveys dramatically increased the number of objects measured, and today it is undoubted that the distribution of galaxies in the Universe is not random but follows a cosmic spider-web like pattern \citep[see, for instance,][for the first quantifications of non-randomness]{Turner1975,Soneira1977}. All galaxy redshift surveys further unveiled a rich tapestry of galaxies, each with unique properties and characteristics, providing valuable insights into the nature of our Universe, its expansion, and the intricate web of cosmic structures. This large-scale structure and cosmic web, respectively, can be readily explained within the framework of the concordance cosmological model that induces hierarchical structure formation \citep[e.g.][]{Zeldovich1970,Frenk1983,Davis1985}. Starting from tiny seed inhomogeneities, gravity amplifies these first matter perturbations eventually leading to the observed distribution of galaxies, galaxy clusters, and the cosmic web in general. However, seed inhomogeneities are needed whose origin is believed to be in the very early Universe, caused by a brief inflationary period during which primordial Gaussian-like fluctuations are generated \citep{Guth1981,Linde1982}.

Since the first surveys and systematic observations of the Local Universe, workers in the field aimed at characterizing and quantifying the distribution of galaxies using various statistical methods, revealing various regular patterns in it. For instance, \citet{Lapparent1986} measured -- starting from the CfA2 catalogue \citep{Huchra1983} -- 584 redshifts from galaxies located in the Coma cluster direction and interpreted this as `bubble-like' structures of a typical diameter of 25 $h^{-1}$Mpc (which these days are rather called cosmic voids). \citet{Tully1978} obtained redshifts for another set of 412 galaxies -- starting from the Palomar Sky Atlas -- finding a peak in the two-point correlation function at a scale of ca. 2.5~Mpc in the Local Universe. Another regularity in the galaxy distribution are the so-called Baryonic Acoustic Oscillations (BAO), i.e. fluctuations in the density of the visible baryonic material, caused by acoustic density waves in the primordial plasma of the early universe \citep[][]{Peebles1970,Bond1984,Holtzman1989}.  The BAO signal shows up as a bump in the two-point correlation function at scale equal to the sound horizon at photon decoupling. While first only theoretically predicted, early hints suggested an observational counterpart (\citet{Percival2001}) which was eventually statistically confirmed by two-point correlation function analysis applied to the SDSS data by \citet{Eisenstein2005} and to 2dF data by \citet{Cole2005}. And there are nowadays even recent claims to actually have found a single such oscillation in the Cosmicflows-4 data \citep[][]{Tully2023_BAO}.

About a decade ago, \citet{McCall2014} reported the existence of a so-called `Council of Giants' (CoG), i.e. a ring-like structure about the Local Group made up of 11 massive galaxies with stellar masses ranging between $M_* \sim 3 \times 10^{10} M_{\odot}$ (M64) and $M_* \sim 9 \times 10^{10} M_{\odot}$ (Maffei 2). Please note that this structure was identified by eye, with no quantification of its stability with respects to variations in distance to and/or luminosity of its member galaxies. However, a related work by \citet{Neuzil2020} confirmed its presence in the Local Volume Galaxy (LVG) catalogue \citep{Karachentsev2013}, an updated version of the catalogue upon which McCall based his study. In their work, they placed the observer at the centre of the Milky Way and found the CoG as a jump in the spherically-averaged, cumulative number density of galaxies. This is a similar approach to \citet{Karachentsev2018} who calculated the (again spherically-averaged) mean density of stellar matter within a distance $D$ in the Local Volume. They also found a peak at $D\sim 3.5$~Mpc where the CoG is located. However, they neither call it a ring-like structure nor make reference to the CoG.

Motivated by the possible existence of the CoG and ring-like structures in general, we have developed a method to find mathematically defined patterns in 3D point distributions such as galaxy catalogues. However, our aim is to make as few assumptions about the location, size, and orientation of such features as possible. We therefore developed a tool that takes as input a 3D distribution of points and automatically searches in it for -- in our case -- toroidal structures, i.e. tori with a certain radius and thickness (i.e. a doughnut-like shape). In doing so, we make no assumptions about the centre-position of these tori. Using this newly developed tool, called HINORA (HIgh-NOise RANdom SAmple Consensus), we study the LVG data of \citet{Karachentsev2013} in search of the CoG and other similar structures for various cuts in K-band luminosity \citep[reminiscant of stellar mass cuts,][]{Bell2003,Karachentsev2018}. The detection of possible rings formed by baryonic (or possibly dark) matter in the nearby small-scale structure could change our understanding of the nature of these components, as well as their behavior. We remark that the HINORA code can universally be applied to any 3D point distribution and hence could also serve to detect aforementioned BAO peak(s). This is possible since this method can be generalized to the search of any simple figure, so its application to spherical shapes would only imply minor geometric changes.

In a follow-up work we will apply the HINORA code to cosmological simulations, both random \citep[such as Illustris,][]{Vogelsberger2014} as well as constrained \citep[such as HESTIA,][]{Libeskind2020}. This will provide insight into the origin of these formations and how common they are in the cosmos. Are such galaxy rings an unknown part of the cosmic web distribution? What if they are merely a coincidence? However, here we restrict ourselves to the description of the HINORA method as well as its application to observational data of the Local Universe.

The outline of the paper is as follows: in Sec.~\ref{sec:data} we will describe the Local Volume Galaxy (LVG) catalogue used for the analysis. In Sec.~\ref{sec:method} we will present the HINORA method capable of recognizing rings or any other regular geometric patterns in point clouds with high noise levels and a substantial degree of background, respectively. In Sec.~\ref{sec:LVG} we will apply the method to the LVG data, presenting the occurance of two distinct ring-like features in our Local Universe. In Sec.~\ref{sec:discussion} we briefly discuss the mode of operation and limitations of HINORA, based upon the results obtained in the previous Section. We close with a summary and the conclusions in Sec.~\ref{sec:conclusions}.

\section{The Local Volume Galaxies Data}
\label{sec:data}
Real-space positions and magnitudes of the galaxies under consideration here are taken from the Local Volume Galaxy Catalog \citep[LVG,][]{Karachentsev2013}.\footnote{\url{https://www.sao.ru/lv/lvgdb/}} These data contain the largest amount of current information on nearby galaxies within the Local Volume. This set is based upon both own observations \citep[see for instance][]{2019AstBu..74..111K} as well as measurements from other sources, and is continuously updated. In it, \citet{Karachentsev2013} compiled measures with a distance up to $D\lesssim 11$ Mpc from the Milky Way, providing the distances of each galaxy, magnitudes in B and K filters, radial velocities, and other information. To deal with the region with the lowest error estimates, we restrict our study to objects found within a sphere of radius 10 Mpc centered on the Milky Way (MW), which leaves us with 1069 galaxies in total. All coordinates have been transformed to supergalactic, having the MW at the centre. Other catalogs of relevance, such as for instance Cosmic Flows (CF, \citet[][CF3]{Tully2016} and \citet[][CF4]{Tully2023_CF4}), cover a much larger volume and provide estimates of radial velocities with respect to the Local Group. However, since we want to cover both the most massive galaxies and the least bright dwarf galaxies, we will prefer to work with the LVG catalogue: LVG provides us with approximately three times as many measured objects in the $\leqslant 10$ Mpc range than CF3 and about 70 per cent more than CF4. 

Further, a remarkable difference between LVG and CF is the considerable systematic discrepancy in the distances provided for some galaxies, such as those belonging to the Maffei Group. As \citet{Neuzil2020} also points out, the distances to Maffei 1 and Maffei 2 are particularly sensitive since they are in the `Zone of Avoidance' of our galaxy and hence subject to redening. In case of the Maffei 2, we took into account recent works by \citet{Tikhonov2018} and \citet{Anand2019}, where the authors discovered that this galaxy lies at a distance of 5.7~Mpc, instead of 3.5 Mpc, as previously thought. The sample of LVG galaxies was taken from the current version of the database of galaxies of the Local Volume (\citet{Kaisina2012}). Inside 6--7 Mpc, the distances to most galaxies have been measured by high-precision photometric methods (Cepheids, RR-Lyres, and the top of the red giant branch) with accuracy of the order of or better than 5 per cent. Distances to more distant galaxies have generally been measured by less accurate methods, such as the Tully-Fisher relation, fundamental plane, the luminosity of the brightest stars, giving the error of more than 20--25 per cent.

Another argument in favour of LVG is that CF provides the luminosity in the B filter, whereas we are interested in K-band magnitudes: the B-band is not a reliable estimator of stellar mass given its sensitivity to galactic internal extinction and its dependence on various inhomogeneously distributed features, such as age, metallicity, or SFH of the nearby galaxies. Those magnitudes in the near-infrared K band in LVG are taken from the 2MASS sky survey \citep{Jarrett2000, Jarrett2003}, supported by measurements by \citet{Fingerhut2010} as well as \citet{Vaduvescu2005,Vaduvescu2006}. However, as already noted by \citet{Kirby2008} and \citet{McCall2014}, even if the 2MASS survey has obtained measurements with a uniform methodology, for certain very low luminosity galaxies it can overestimate the K magnitude by up to 2.5 mag due to short integration time, while for bright galaxies the luminosity may not be correctly measured given the finite extrapolation of the radii. This uncertainty will be important, and we will take it into account by reabsorbing it into the error in the distance measurement. If there is no data available for the K-band of one of the objects, the authors \citet{Karachentsev2013} estimate its value using methods that rely on other measured bands, as detailed in their paper. The K-band is a reliable stellar mass $M_*\sim (M_\odot/L_\odot) L_K$ indicator that follows a stellar $M/L$ relation with less color dependence than the other bands \citep{Bell2001,Bell2003,Beare2019}. While in the K-band, the $M/L$ ratios for spiral galaxies in the study of \citet{Bell2001} vary by as much as a factor of 2, in the B-band, they vary by a factor of 7. The work by \citet{Bell2003} updates these estimates using SDSS and 2MASS photometry and provides $M/L_{K}$ ratios. Whenever needed or of interest, we will simply convert the K magnitude to stellar mass assuming $M/L_{K}\sim 1$, as done by the authors of LVG \citep{Karachentsev2013}. 

\section{The Method}
\label{sec:method}
Our prime objective is to find and quantify ring-like structures in a point distribution (i.e. in our case the LVG catalogue). We need to approach the problem through a systematic process that quantifies and confirms the presence of patterns within a system composed of a spatial distribution of discrete points. The main techniques for the systematic search of basic patterns in 3D point distributions are the Point Cloud Segmentation (PCS) algorithms \citep{Xie2019}. PCS is based on the use of simple geometric rules to find 2 or 3 dimensional structures with low noise levels. It is an unsupervised method, so the algorithm locates system relationships by analyzing the data without the need for an external instructor during the process, and without a previous sample dataset. The PCS family in turn contains several independent algorithms that have been developed in recent years \citep{Xie2019}. One of them is RANdom SAmple Consensus (RANSAC), an algorithm based on testing models that fit a randomly chosen subset of the data \citep{Fischler1981}. Each model is evaluated by calculating how many points of the total set are adequately approximated by it. After multiple steps, the algorithm chooses the model that contains the largest number of points. RANSAC is the basis of the methodology chosen in this work, and its way of operation will be explained in more detail below.

\subsection{The RANSAC algorithm and its variants}
\label{sec:RANSAC}
RANSAC is a randomized pattern search algorithm applied to point clouds \citep{Choi2009,Raguram2008,Raguram2013}. For its application, it is necessary to define the pattern to be searched and the size of the model $\tau$. The distance $\tau$ will characterize the maximum distance that a point of the figure is allowed to have for it to still be considered part of it. The pattern is defined by basic geometry from the starting subset. If we were looking for lines, the subset would have to have 2 points. RANSAC would pick 2 random points from the data, and join them with a line. All the points at smaller distance than $\tau$ from that line would be considered as inliers, and would form part of the model. The model is therefore composed of the points closest to the geometric line formed by the random points matched by RANSAC. In our case, we work with circles: to draw a circle in space, at least 3 points are needed. Therefore, the subset that RANSAC collects will be composed of 3 random points of the data, for which it will draw a circle that joins them. The rest of the points are checked for possible inclusion in the model via distance $\tau$ to this ring. As a consequence our model points will lie inside a torus. RANSAC operates in the following two phases:

\paragraph*{Phase 1:} In the first phase, RANSAC generates a hypothesis of the model. A small subgroup is formed from randomly selected points of the data, and these points will be used to geometrically draw the figure in space. This figure will be called a ``model''. In our case it consists of 3 points at this stage.

\paragraph*{Phase 2:} The second phase consist in evaluating the hypothesis generated by the random subgroup. The minimum distance to the figure of all the points belonging to the full cloud is calculated. Those points that are at a distance less than $\tau$ from the model are considered to belong to it, and will be called inliers. All other points of the cloud will be considered outliers.

If the number of inliers is higher than that calculated for previous hypotheses, both the coordinates of the generated figure and its inliers are saved and the process is started again form Phase 1, trying to find again the ring with more inliers. The final ``correct'' model will be the one that contains the highest number of inliers after numerous iterations of the code. Equation~(\ref{ec:methodec1}) gives the number of iterations $N$ required to find a particular model in a point cloud with a probability of success $p$ \citep{Choi2009}

\begin{equation}
N=\frac{\log(1-p)}{\log(1-(1-e)^s)} \,
\label{ec:methodec1}
\end{equation}

\noindent
where $s$ is the number of points forming the sample subset, and $e$ is the proportion of system outliers. In our case $s=3$ (as we start with 3 points in the Phase 1). We further set the desired success rate to $p=0.99$ and $e=0.85$. In our case, higher values of $p$ do not alter the results so even if the algorithm does not converge exactly it is not necessary to increase this number. On the other hand, the value of $e$ allows us to find patterns with inliers of at least $15\%$ of the data, which will be important as we will explain in Sec.~\ref{sec:LVG}. Substituting these numbers into Eq.~\ref{ec:methodec1} reveals that we need of order 1400 iterations of Phase~1+Phase~2 before analysing the possibly found rings. We also need to mention that several runs of RANSAC on the same data may return the same pattern but with slightly different characteristics if it has been formed by a different subset of inliers. We will take advantage of this to enable our method to locate a certain ring more precisely by taking advantage of these multiple detections.

RANSAC has several variants that modify the original algorithm in favor of enhancing certain features \citep{Choi2009}. The Randomized RANSAC (R-RANSAC) technique used to reduce the computation time consists of a prior evaluation of the hypothesis. R-RANSAC introduces a preliminary test of the data before evaluating it completely in each iteration \citep{Matas2004}, which will allow discarding rings generated > 10 Mpc from the Milky Way. Further, to adapt RANSAC to search for faint figures in data with noise levels greater than 90$\%$ of the total points, we will make modifications to the original algorithm, to be explained now. 

\subsection{High Noise RANSAC (HINORA): adaptation of the algorithm  to small inliers ratios}
\label{sec:RANSACmod}
\begin{figure}
	\includegraphics[width=\columnwidth]{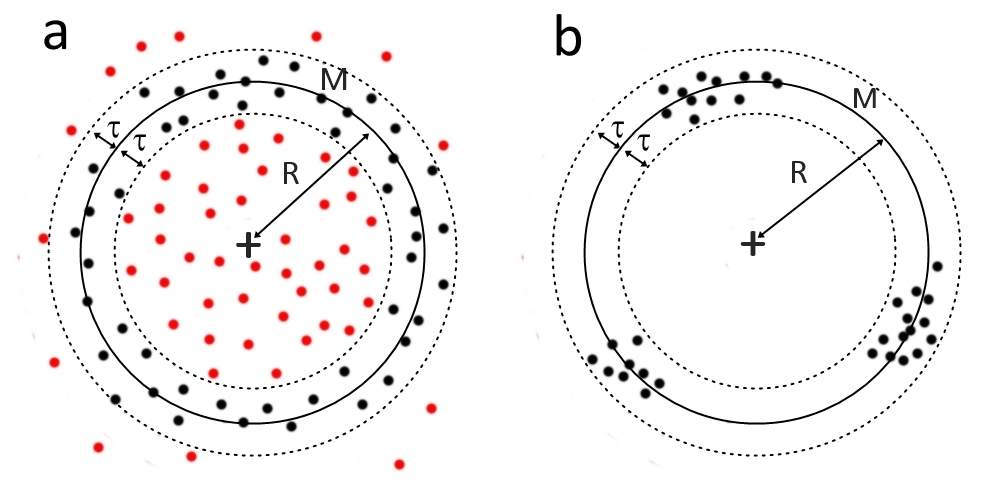}
    \caption{2D examples of the failures of RANSAC when applied to noisy data. Panel a represents how the method mistakes a big noise accumulation for a valid ring. Panel b shows how the algorithm mistakes small noise clusters for a valid ring. The cross marks the center of the ring. The black solid line represents the model M, located at a distance R from the center and with an inner radius $\tau$ bounded by the dashed lines. The black and red dots correspond to inliers and outliers respectively.}
    \label{fig:2problems}
\end{figure}

In order for RANSAC to give credible results for the data at hand, we need make some adjustments. The reason for that is illustrated in Fig.~\ref{fig:2problems}. There two situations are depicted for which RANSAC will return a positively found ring-like structures:

\begin{itemize}
    \item[a)] A random distribution of points can be recognized as a valid model just because it contains a large amount of data.
    \item[b)] Clusters of points, distributed anisotropically on the circumference of a ring match the pattern constraints, too.
\end{itemize}

Regarding point a), the original RANSAC code seeks to maximize points that are inside the model, without considering anything else. However, this is fatal when dealing with noisy environments, because RANSAC will only look for sites with many points -- whether they are noisy or not. For instance, if the data contains 90 per cent noise, the original RANSAC will tell you that the model is where it finds the most points. In other words, the original RANSAC code identifies overdensities following the shape of the desired model. By introducing the $\alpha$ parameter (see Sec.~\ref{sec:alpha} below), we punish the algorithm when it finds only large overdensities, and force it to focus only on overdensities that have the shape of the model and that have an environment without many points outside the ring.

Both situations are problematic since clearly neither of them corresponds to a hypothesis that should be approved. In the first case, the algorithm does not distinguish rings and accumulations of symmetric noise distribution. Structures such as spheres or simple concentrations of random data can be mistaken for rings since the algorithm does not take into account the shape of the outliers in the near environment of the model (panel \hyperref[fig:2problems]{a}). RANSAC will only look for sites with many points -- whether they are noisy or not. The original version of RANSAC seeks to maximize points that are inside the model, which could be interpreted as searching for overdensities in the data. But in our case we also care about the embedding within the environment, aiming at the lowest possible number of outliers about the identified structure. In the second case, the algorithm does not take into account the distribution of inliers. Although the points are concentrated without noise around them, for RANSAC they can form rings from concentrated clusters of points. This is again a situation we try to avoid.

In order to address these issues, we made modifications to the evaluation step in RANSAC. The most important quantity for each model will be the number of inliers, as this will decide its relevance in the total data. However, this information is not sufficient to solve the problems seen in Fig.~\ref{fig:2problems}, and therefore we will below define two new parameters calculated at each iteration that passes the original hypothesis test of RANSAC. The first one will be the the noise level (the relationship between the model and the points of the environment that are not part of it); and the second one will be the level of regularity that the inliers have (the relationship that the points of the model have with each other). Note that these parameters are provided for each successful ring returned by RANSAC and only evaluated in post-processing.

\subsubsection{Quantification of data noise}
\label{sec:alpha}
\begin{figure}
	\includegraphics[width=\columnwidth]{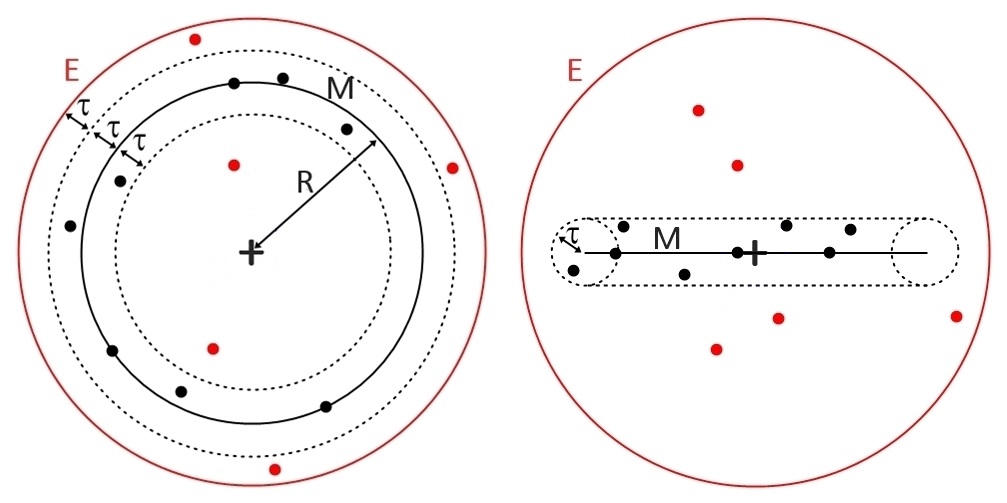}
    \caption{3D example of the defined point types used during the calculation of the data noise parameter $\alpha$. The image on the left represents a projection in the plane of the ring (i.e. model $M$ that forms a torus), and the image on the right an edge on view. The plus sign marks the center of the ring. The red solid line denotes the range of the environment $E$, and the black solid line the model $M$ with $R\pm\tau$ being the outer and inner radius of that torus, respectively. The dashed lines represent the range of the model, distanced $\tau$ from $M$. The black and red dots correspond to ``inliers'' and ``outliers'', respectively. Since the hypothesis has 8 inliers and 5 outliers, the value of $\alpha$ is 0.38.}
    \label{fig:alpha}
\end{figure}

The first parameter we define will categorize all points within an environment $E$, larger than and hence encompassing the model $M$ that can be considered a torus. This will relate points that are part of an existing pattern to nearby, but not relevant points. The volume $E$ will be that of a sphere that shares its center with the circular pattern and will have a radius of $R + 2\tau$. The motivation for this maximal extends characterized by $2\tau$ is to stay within the local environment $E$ of the ring $R \pm \tau$. The points within that sphere $E$ will be categorized as either ``inliers'' (points within the model, denoted by the letter ``$I$'') or ``outliers'' (points within the environment but outside the model, denoted by the letter ``$O$''). Note, while our ``inliers'' are identical to the inliers defined by RANSAC, our ``outliers'' are limited to the volume defined by $E$. Fig.~\ref{fig:alpha} shows an example of how these categories are distributed in environment $E$ (red circle): All inliers are at a distance less than $\tau$ from $M$, and are represented as black dots; the remaining points of $E$ are outliers, denoted by red dots.

With these considerations, we define the data noise parameter
\begin{equation}
\alpha=\frac{N_O}{N_I + N_O}
\label{ec:methodec2}
\end{equation}
where $N_I$ ($N_O$) denotes the number of ``inliers'' (``outliers''). Note, both $N_I$ and $N_O$ are functions of $\tau$ whose value we will motivate later on in Sec.~\ref{sec:LVG}. By means of the parameter $\alpha$ we have related the hypothesis to its environment and we can distinguish situations such as those seen (in the left side of) Fig.~\ref{fig:2problems}. This quantity varies between 0 and 1. If $\alpha$ is small, most of the points belong to the model and the hypothesis is of quality. If $\alpha$ is close to 1 it will be difficult to detect a ring. However, $\alpha\rightarrow 1$ does not always indicate the non-existence of rings, but it does rule out that they have a statistically more relevant presence in the studied environment than in the rest of the data. To be considered, the models have to overcome the local noise.

\begin{figure}
	\includegraphics[width=\columnwidth]{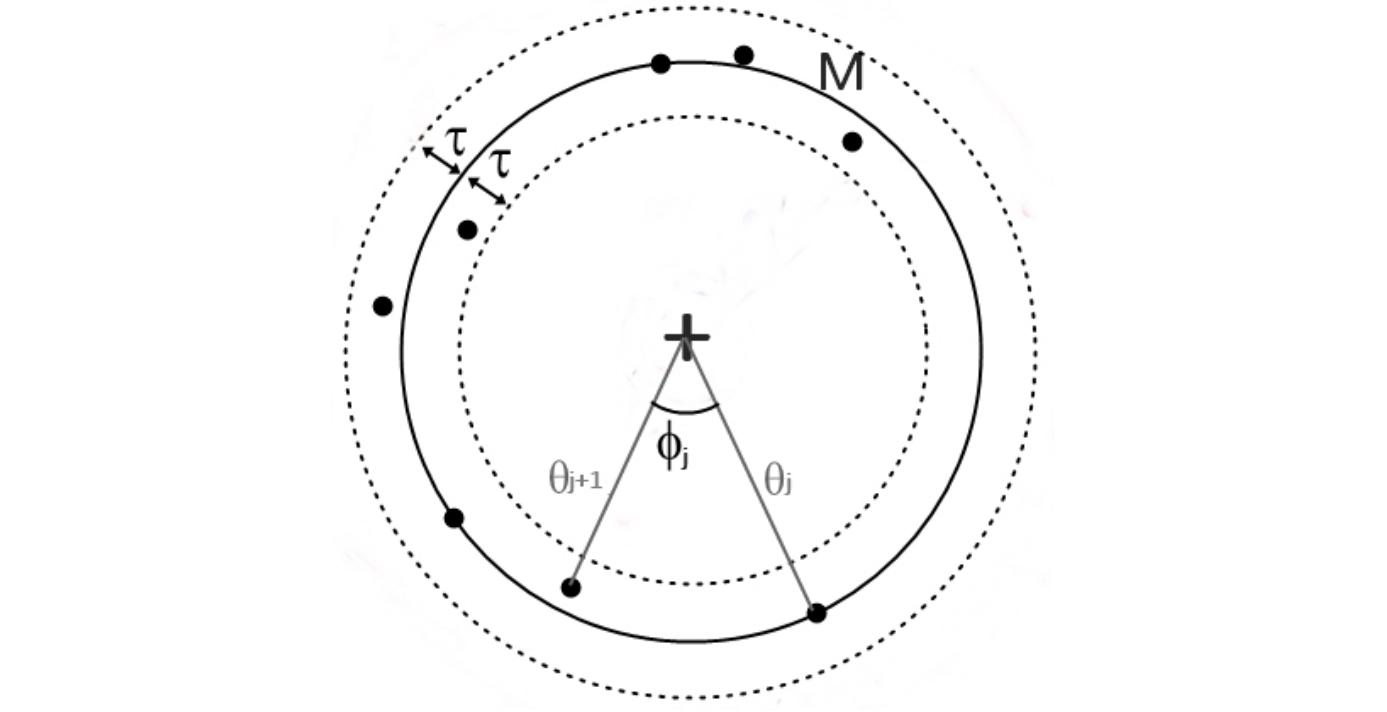} 
    \caption{Example of the projection of a ring on its plane, on which the angular coordinate of the sample inliers $\theta_j$  and the angular separation $\phi_j$ are shown. The value of $\beta$ here is 0.69.}
    \label{fig:beta}
\end{figure}

\subsubsection{Quantification of data regularity}
\label{sec:beta}
To quantify the isotropy of the points, we need a parameter that quantifies how evenly the angles of the inliers are distributed in the model. All the inliers are projected onto the plane in which the hypothesis torus is located. Subsequently, each inlier is expressed in polar coordinates, giving us the azimuthal angle $\theta_i$ with $i\in1,2,...,N_I$. We determine $\phi_j$ as $\phi_j=\theta_{j+1}-\theta_j$ with $j\in1,2,...,N_I-1$, satisfying that $\sum_{j=1}^{N_I-1}\phi_j=2\pi$ (exemplified in Figure \ref{fig:beta}). We then quantify the isotropy as
 \begin{equation}
\beta=\frac{\sigma_\phi}{\langle{\phi}\rangle} \,
\label{ec:methodec3}
\end{equation}
with $\sigma_\phi$ the standard deviation of the $\phi_j$, and $\langle{\phi}\rangle$ their mean. $\beta$ measures how different the $\phi_j$ are from each other by normalizing to the mean of this quantity so that it becomes independent of the number of inliers (note that in general the more inliers, the smaller the values of $\phi_j$). Although this is not the only spherical coordinate involved in the pattern, the quality of the ring does neither depend on the standard deviation of the polar angle nor on the radius. However, it is critical not to approve irregularities in the angle $\theta_i$ such as those in panel \hyperref[fig:2problems]{b} of Figure \ref{fig:2problems}. If the term $\beta$ is small it means that there is no variety in the angular separation of the data and therefore the distribution forms a uniform ring. Or put differently, values $\beta >> 1$ entail that the variation of angles is larger than the mean angle itself, which should not happen. 

\subsection{Validity margin estimation}
\label{sec:vme}
During each evaluation of a hypothesis as defined by a random subgroup (Phase 2 of RANSAC, see above), the three values $N_I$, $\alpha$ and $\beta$ will be calculated for that putative ring. The hypothesis will be accepted, if these three parameters exceed a certain limit $\bar{N_I}$, $\bar{\alpha}$ and $\bar{\beta}$ defined for the respective data set.Therefore, a model will only be considered valid, if it satisfies $N_I\geq\bar{N_I}$, $\alpha\leq\bar{\alpha}$ and $\beta\leq\bar{\beta}$ (models with values worse than the established $\bar{N_I}$, $\bar{\alpha}$ and $\bar{\beta}$ values are considered as spurious objects caused by noise). $\bar{N_I}$ is completely determined as soon as we decide the maximum fraction of outliers $e$, since it satisfies $\bar{N_I} = (1-e) N_{tot}$ where $N_{tot}$ is the total number of points that the data contains. The maximum generalization of this method is always sought, which implies that values for $\bar{\alpha}$ and $\bar{\beta}$ need to be estimated by procedures independent of the input data, thus turning the algorithm into a black box. If that can be achieved, it is only necessary to set the model size $\tau$ and the minimum number of desired inliers $\bar{N_I}$. Note that $R$ is a free parameter that is allowed to vary within a reasonable range: it does not make sense to have $R\leq\tau$ or $R$ greater than the extent of the 3D data.

\paragraph*{$\bar{\alpha}$ margin estimation.} To estimate $\bar{\alpha}$ we will calculate the expected number of noise points in a given data environment $E$. Let $N$ be the total number of points in a volume $V_T$. If our hypothesis now conatins ${N_I}$ inliers, that total space will contain $N-N_I$ points (that are all outliers of the model). Recall that in each if our refined evaluation a spherical environment $E$ of radius $R_E=R+2\tau$ is constructed, with $R$ being the radius of the ring. Therefore, the expected number of outliers in $E$ would be $N-N_I$ multiplied by the fraction of the total volume $V_T$ occupied by the environment $E$. Combining these expressions with Eq.~\ref{ec:methodec2}, we obtain:

\begin{equation}
\bar{\alpha}=\frac{(N-{N_I})\frac{V_E}{V_T}}{{N_I} + (N-{N_I})\frac{V_E}{V_T}} \,
\label{ec:methodec4}
\end{equation}
where $V_E=\frac{4}{3}\pi R_E^{3}$ is the volume of environment $E$ defined for a certain ring. This will be the reference value against which we are comparing the actual $\alpha$ value for a given hypothesis.

\paragraph*{$\bar{\beta}$ margin estimation.} To estimate $\bar{\beta}$ we will calculate the standard deviation and the mean of the distribution of all the 3D points that make up the data. Let $\vec{r_{i}}$ be the position of each point belonging to the data, with $i=1,2,...,N$. The position of the centroid of the point cloud is calculated by $\vec{r_C}=\frac{\sum_{i=1}^{N}\vec{r_{i}}}{N}$. Subsequently, the distance $D_i$ from each point $\vec{r_{i}}$ to the centroid $\vec{r_C}$ is computed. Defining $d_j$ as $d_j=D_{j+1}-D_j$ with $j\in1,2,...,N-1$ (the $D_j$ terms are ordered from smallest to largest), we can calculate $\bar{\beta}$ as follows:
\begin{equation}
\bar{\beta}=\frac{\sigma_d}{\langle{d}\rangle} \,
\label{ec:methodec5}
\end{equation}
with $\sigma_d$ the standard deviation of $d$, and $\langle{d}\rangle$ its mean. The term $\langle{d}\rangle$ in Eq.~\ref{ec:methodec5} has a similar function to $\langle{\phi}\rangle$ in Eq.~\ref{ec:methodec3}. On the one hand it makes the parameter dimensionless, and on the other hand it makes it independent of the number of points. One of the reasons for choosing this methodology among others to find the standard deviation of the point cloud is because for 
data where all points form a symmetric ring, $\bar{\beta}$ tends to be 0.\\

\begin{figure}
	\includegraphics[width=\columnwidth]{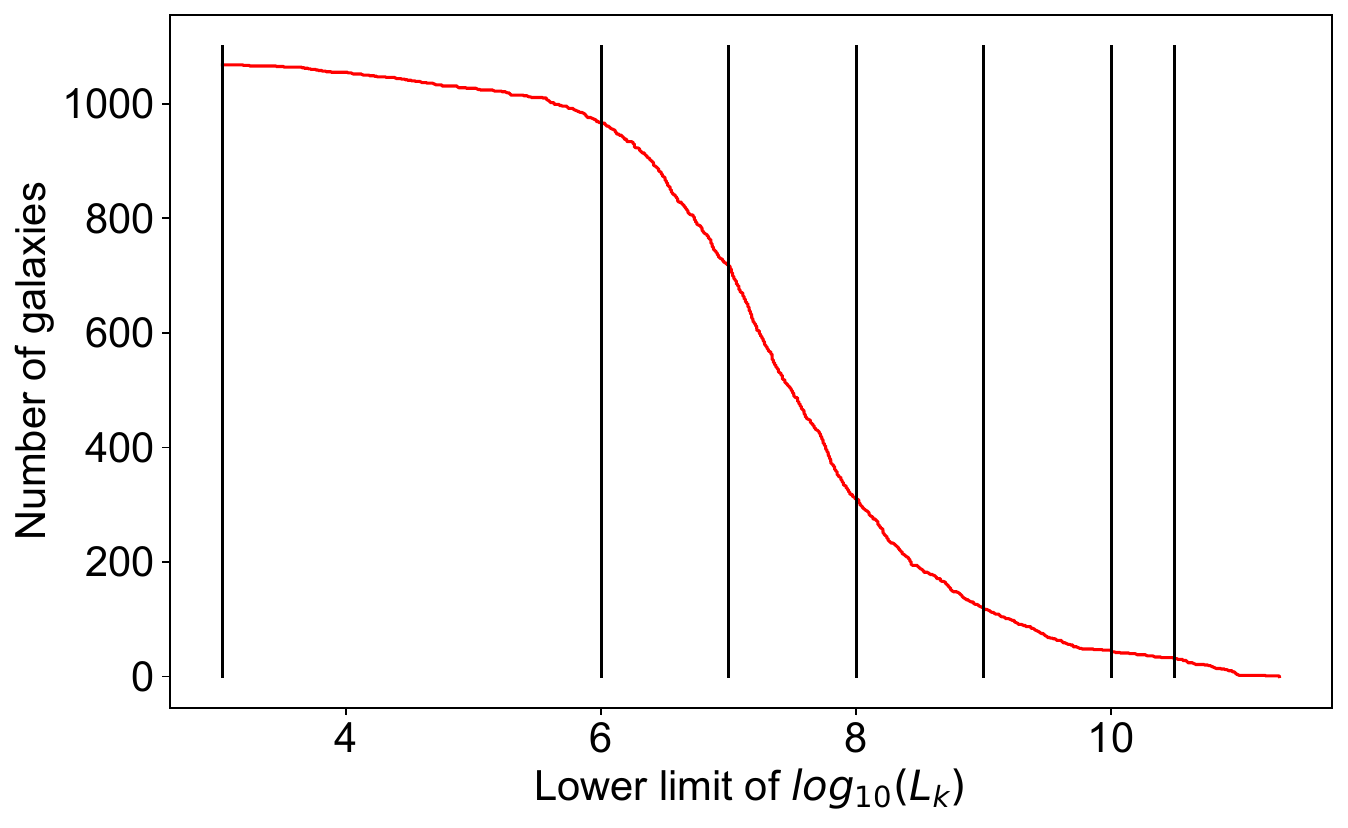} 
    \caption{Cumulative luminosity of the LVG catalog limited to nearby galaxies with a distance < 10 Mpc to the MW. The vertical lines represent the luminosity cuts applied to the data, i.e. $\log_{10}L_{K}$ = 3, 6, 7, 8, 9, 10, and 10.5.}
    \label{fig:luminosity}
\end{figure}

For brevity, in the following we will refer to this method as ``HINORA'' (HIgh NOise RANSAC). The most important input parameter to HINORA is the size of the pattern $\tau$. Upon exit, HINORA provides us with the following information, in case a ring has been successfully found: the position of the ring centre, the radius of the ring,\footnote{While $\tau$ is a fixed input value, HINORA determines the credible range of possible ring radii itself by searching from $R_{\rm min}=2\tau$ to some user-defined maximum radius $R_{\rm max}$ (in our case 10 Mpc), always ensuring that any probable ring fully lies within the complete data range.} the normal vector to the plane of the ring, and our newly defined quality assessments $\alpha$ and $\beta$. As mentioned before, the algorithm can find the same structure multiple times, but with marginally different inliers. This is corrected for by identifying all rings that share 30 per cent of inliers and then only keeping the one with the best values of $N_I$, $\alpha$ and $\beta$.\\

Before eventually applying HINORA to the observed galaxy catalogue LVG, we have subjected the newly designed code to a series of credibility tests. The results of these validations are summarized in \ref{tests}. Those final assessments have revealed that HINORA reliably identifies ring-like structures, if they are really present.

\begin{figure*}
    \centering 
	\includegraphics[width=0.8\linewidth,trim={3cm 1.5cm 3cm 2.5cm},clip]{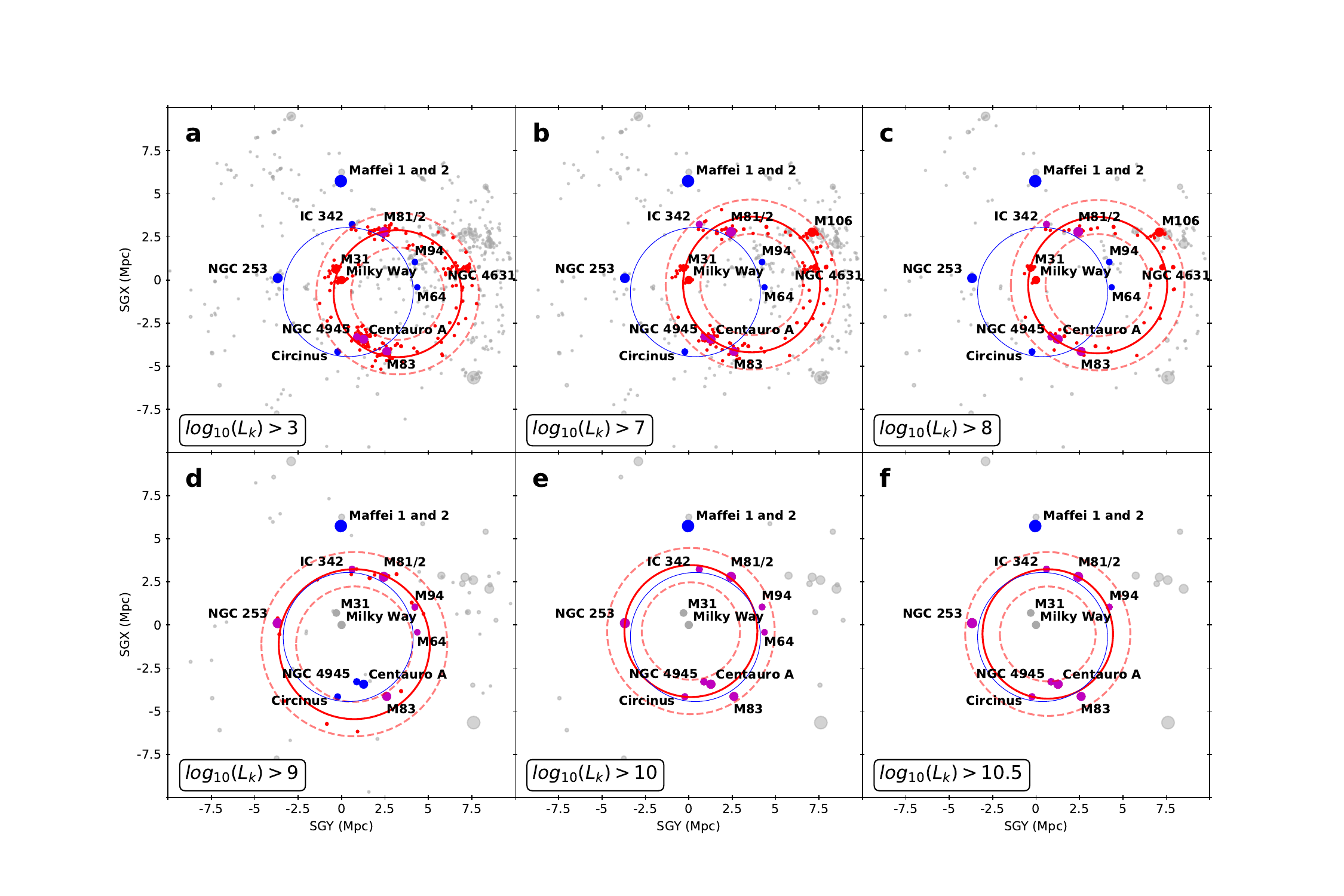} 
    \caption{Projection along the $z$-axis of LVG galaxies (in supergalactic coordinates) for four K-band luminosity cuts $L_K=[3,7,8]$ (upper panel) and $L_K=[9,10,10.5]$ (lower panel). The size of the dot is proportional to $\log_{10}(L_K)$. In red and magenta we show galaxies belonging to the ring as identified by HINORA, where the ring and its respective $\tau$ are also indicated by solid and dashed red lines, respectively. The original galaxies making up McCall's Council of Giants are represented as blue and magenta points, also highlighting the corresponding ring as a solid blue line. Note, magenta points belong to both rings.}
    \label{fig:map}
\end{figure*}

\section{Ring-like structures in LVG}
\label{sec:LVG}
The main motivation for this work (i.e. the development of an automated finder for pre-defined structures in 3D point distributions) comes from the observations first put forward by \citet{McCall2014} who reported the existence of a so-called Council of Giants, i.e. ring-like structure about the Local Group made up of 11 massive galaxies with stellar masses ranging between $M_*=10^{10.496} M_{\odot}$ ( M64) and $M_*=10^{10.928} M_{\odot}$ (Maffei 2). However, as argued above in Sec.~\ref{sec:data}, we prefer to work with the to-date most complete survey of galaxies in the Local Universe, i.e. the LVG catalogue \citep{Karachentsev2013}. By applying HINORA to it, we have higher statistical reliability and use more current measurements of, for instance, distance (which is crucial to our objective). These data contain about a thousand galaxies with both distance and K-band luminosity $L_K$ measurements. When translating $L_K$ into stellar mass, we will assume a bilateral linear relationship. In this work we are not going to go beyond such a simple approximation which is sufficient to indicate whether or not the galaxies can be considered massive \citep[see, for instance,][]{Jarrett2013,Ziparo2016}. And in order to find any possible mass trends in our ring-finding, we are going to apply several cuts in K-band magnitude. To find the most suitable lower $L_K$ limits, we show in Fig.~\ref{fig:luminosity} the cumulative K-band luminosity function. The first cut applied by us is at the luminosity of the least bright LVG object ($\log_{10}L_{K} = 3.03$)\footnote{We will for simplicity use and denote this cut as $\log_{10}L_{K} = 3$} and thus includes 100 per cent of the galaxies in this catalog. The following cuts are simply taken at $\log_{10}L_{K} = 6, 7, 8, 9, 10$, and $10.5$ corresponding to about 90, 65, 30, 10, 4, and 3 per cent of the total objects, giving us successively more massive galaxies. Each of the seven (sub-)sets has been passed to HINORA for the possible detection of ring-like structures. For that we set the size $\tau$ of the pattern to $\tau=1$~Mpc, and the number of minimum inliers $\bar{N_I}$ to $\bar{N_I} = 0.15 N_{tot}$. Note that the choice of $\tau$ is driven by our motivation to detect ring-like structures in the LVG data that have radii larger than 1~Mpc yet still lie within the domain of the catalogue. We also varied $\tau$ in-between 0.5~Mpc and 2~Mpc, counting the number of rings found for each of our applied luminosity cuts. Though not explicitly shown here, we found that for $\tau=1$~Mpc we always do find a ring while for larger and smaller values the likelihood for it decreases towards zero. Note that another value we have set is $\bar{N_I}$. Our choice of $\bar{N_I} = 0.15 N_{tot}$  is motivated because we wish to capture the possible existence of the CoG, which represents $\sim 0.2 N_{tot}$ in the \citet{McCall2014} catalogue. Using values higher than $0.2 N_{tot}$ would prevent us from finding the possible CoG or rings similar to it, while lower values would capture false models coming from noise. For example, for the highest luminosity cuts, choosing $\bar{N_I} = 0.1 N_{tot} \approx 3$ causes the algorithm to find correct rings consisting of only 3 points. $\tau$ and $\bar{N_I}$ are the only parameters to be chosen beforehand, since the radius of the ring and all other return values (cf. Sec.~\ref{sec:RANSAC}) are determined by HINORA itself.. And as mentioned before in Sec.~\ref{sec:data}, we are only working with galaxies with a maximum distance of 10 Mpc to the MW.\\ 

\begin{figure*}
    \centering 
	\includegraphics[width=0.8\linewidth,trim={3cm 1.5cm 3cm 2.5cm},clip]{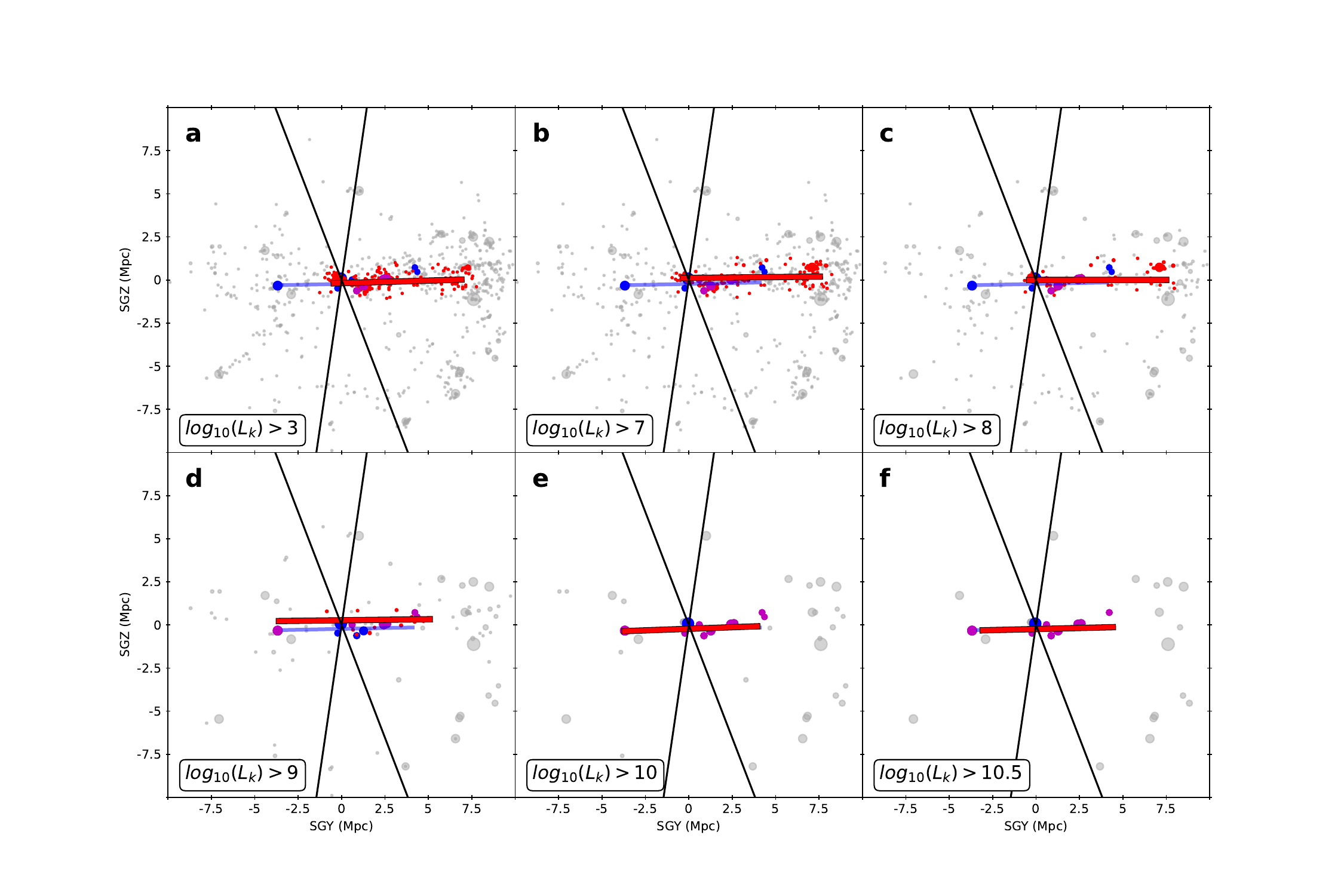} 
    \caption{Same as Fig.~\ref{fig:map}, but now projecting along the super-galactic $x$-axis. The two solid lines delineate the Zone of Avoidance (|b|$\leqslant 10^{\circ}$).}
    \label{fig:mapz}
\end{figure*}

\begin{figure}
	\includegraphics[width=\columnwidth]{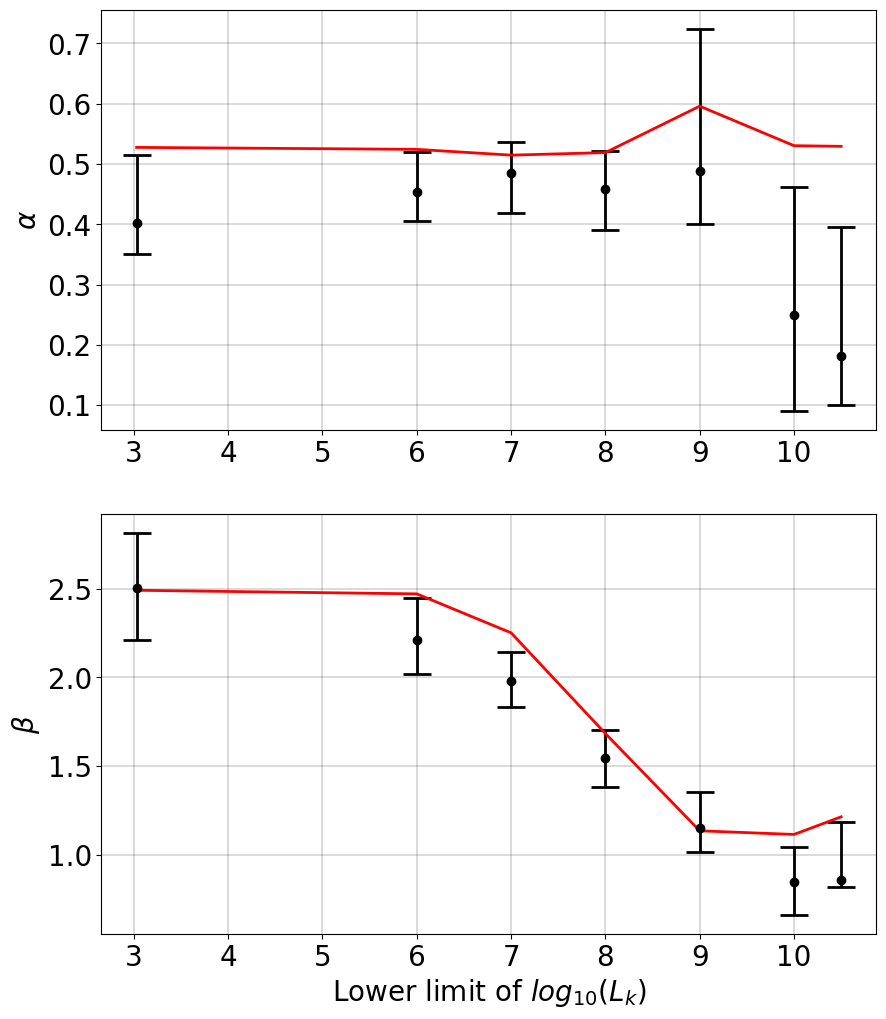} 
    \caption{The median of the $\alpha$ and $\beta$ parameters (explained in sections \ref{sec:alpha} and \ref{sec:beta} respectively) of the models found by HINORA for each LVG luminosity cut-off. The red line represents the values of $\bar{\alpha}$ and $\bar{\beta}$, which define the maximum of $\alpha$ and $\beta$ in the LVG catalog.}
    \label{fig:modelsalphabeta}
\end{figure}

The results can be viewed in Figs.~\ref{fig:map} (face-on view in super-galactic coordinates) and \ref{fig:mapz} (edge on view in super-galactic coordinates), however, we prefer to simply focus and discuss the former face-on representations. In these plots we show the ring found by HINORA (red solid line), the applied $\tau$ range (red dashed lines), and additionally the Council of Giant as reported by \citet[][blue solid line]{McCall2014}. The galaxies belonging to the rings are colour-coded, too. Note that galaxies belonging to both rings are shown in magenta. There are obviously several interesting points to notice in these plots. First and foremost, HINORA always found a ring in the LVG data; and it only ever found a single ring. This structure appears to be rather stable (see stability discussion below), but we actually have to distinguish two scenarios here: one ring up to luminosity cut $\log_{10}(L_K)<9$, and yet another one for $\log_{10}(L_K)>9$. The former one even includes the MW and M31, and all its satellites. In fact, that particular structure is predominantly defined via satellites in general and not their (massive) host galaxies, something that we will later see is reflected in the quality parameter $\beta$ (i.e. the one that measures isotropy) of the two rings. We will refer to this structure also as `satellite ring'. Only when taking into account the most massive galaxies with $\log_{10}(L_K)>9$ we identify a structure akin to McCall's Council of Giants. 
In fact, for galaxies $\log_{10}(L_K)\geq 10$ we do recover the orginal Council of Giants, though excluding Maffei 1 and 2 due to their different distances in the LVG catalogue \citep[][]{Karachentsev2013}. This finding now confirms two things: a) the massive galaxies in the Local Volume do form a Council of Giants, and b) there appears to be a another (ring-like/circular) structure associated with a selection of host and their satellite galaxies. However, it remains unclear why both rings show comparable sizes of approximately 4 Mpc in radius, yet having their centres in different places. 

Fig.~\ref{fig:mapz}, which shows the same data but viewed edge-on, simply confirms that both these structures are planar rings, lying within the Local Sheet. The Local Sheet is part of a larger flat structure, the Local Supercluster with the center in the Virgo cluster (\citet{Tully2008}). Immediately above it in supergalactic coordinates the Local Void begins. On the opposite side, parallel to the plane of the Local Supercluster lie the Leo Spur and Dorado cloud (\citet{Tully1987}). The so-called Avoidance Zone (ZoA) in the Milky Way is shown by solid lines. ZoA cuts the Local Sheet into two halves. Fortunately, due to its orientation, it covers a quite small portion of the Local Sheet and should not significantly affect the detection of the ring structures. Nevertheless, it should be emphasized that due to the extremely strong extinction near the Milky Way plane, the detection of galaxies, as well as the determination of their parameters, becomes an extremely difficult task. In our case, this manifested itself in the erroneous distances to the Maffei 1 and 2 galaxies adopted in earlier works. In turn, this is the main reason for the differences between the CoGs detected by us and those found in \citet{McCall2014}. Further, in principle one should adjust HINORA to allow for such incomplete data sets. But we also acknowledge that the ZoA does not significantly affect the measure of the distribution of positions in the LVG catalogue. This is because Eq.~\ref{ec:methodec5} works with the distances from the centroid to the position of the rest of the data. And since the LVG catalogue is centred on the MW, the centroid will tend to be located close to our galaxy and therefore the ZoA will have no effect on the calculation of these distances. However, in other cases (such as working only with areas far from the MW) and when working with less uniform data, we might have to change our strategy and use some kind of privileged axis along which to project the points and calculate distances.

We like to mention that we have repeated the ring-finding, but assigning a weight to each data point by accordingly adjusting Eqs.~\ref{ec:methodec2} and \ref{ec:methodec4}. We have used both the K-band luminosity $L_K$ as well as its logarithm $\log_{10}L_K$ as weight. While not explicitly shown here, we find that in both cases the Council of Giants ring (i.e. the lower panel of Fig.~\ref{fig:map}) shifts to include both Maffei 1 and 2 (excluding IC342, M81/2, M94, and M64). This same ring -- which obviously has a much lower isotropy level -- is also found for the lower luminosity cuts (i.e. the upper panel of Fig.~\ref{fig:map}) when using linear K-band weights, indicating that Maffei 1 clearly dominates the sample when taken its K-band luminosity (and hence stellar mass) linearly into account. Finally, using the logarithm of the K-band luminosity as a weight, the so-called satellite ring is not affected.

One might now question the importance of these rings as their manifestation depends on the way they are searched for. Answering this is actually beyond the scope of this particular work, in which we primarily introduce the ring-finding algorithm, and showcasing its performance by applying it to astrophysical data. But we nevertheless like to share our thoughts here. While \citet{McCall2014} reported the existence of the Council of Giants simply by visual inspection, we now confirm that this structure can be found, even when using an unbiased and automated method. But while investigating possible cuts and weighing schemes using a proxy for stellar mass (remember, the peculiarity of the Council is the fact that it is made up of `Giants') we found that this structure is not necessarily unique. This is primarily driven by Maffei 1, a giant elliptical galaxy in the Zone of Avoidance whose distance is not the best established one: when basically forcing HINORA to include it by using linear K-band weights, other lower-mass galaxies are removed from the non-weighted ring. We believe that, given the special situation for Maffei 1 (and 2), the results shown here in Fig.~\ref{fig:map} should be the ones to interpret, which is also why we defer from showing the results for the different weighing schemes. Note that both Maffei 1 and 2 do lie in the Local Sheet, which is why HINORA picks them up as inliers in the rings found by it. Maybe the real scientific question here is why there is a Local Sheet of galaxies, with massive galaxies arranged in such peculiar ring-like way. We will address these problem in a follow-up paper where we make use of simulations of cosmic structure formation to gauge the likehood of the existence of such galaxy arrangements. To that extent both constrained Local Universe as well as standard cosmological simulations shall be applied.

Before quantifying the stability of the two rings seen in the previous figures, we will first show the quality assessment of them. To that extent we show in Fig.~\ref{fig:modelsalphabeta} the relation between $\alpha$ (upper panel) and $\beta$ (lower panel), respectively, and the luminosity cut. We have also drawn in each case $\bar{\alpha}$ and $\bar{\beta}$ with a red line, emphasizing the overall behavior of the data for these definitions. We observe that while $\alpha$ has a constant value of less than unity, $\beta$ and $\bar{\beta}$ drops substantially towards $\beta\approx1$ as massive galaxies become more and more important. $\beta$ is particularly small for the Council of Giants, confirming its isotropy. Since in principle $\beta$ and $\bar{\beta}$ are a normalized parameters with respect to the number of objects, they does not depend on the number of inliers, but on their distribution. And we have qualitatively seen in Fig.~\ref{fig:map} that non-uniformly distributed satellite galaxies substantially contribute to the inliers for low $\log_{10}(L_K)$ values: they are clustered about their hosts, however, these hosts are not necessarily the ones that eventually form the Council of Giants. We further note that the actual Council of Giant ring has a substantially lower $\alpha$ than $\bar{\alpha}$, indicative of a more stable structure.\\

We need to remark that Fig.~\ref{fig:modelsalphabeta} features error bars attached to the measured values of $\alpha$ and $\beta$. These have been derived the following way. The LVG catalogue not only lists the distance to a galaxy, but also an error on that estimate. We have now generated thousand `mock LVG' catalogues in which we varied the distance of each galaxy as

\begin{equation}
    D_{\rm LVG} \rightarrow D_{\rm LVG} + norm(\sigma_D) ,      
\end{equation}
where $norm(x)$ is a normal distribution centered at $0$ with standard deviation $x$, and $\sigma_D$ the distance error as given in the LVG catalogue.\footnote{Note that the error on the distances $\sigma_D$ is smaller than the distance $D_{\rm LVG}$ itself.} HINORA had been applied to each of these mock catalogues and the dots shown are the median, and the error bars contain $85\%$ of the generated models. This causes that in Fig.~\ref{fig:modelsalphabeta} we can see $\alpha$ and $\beta$ values above their defined limits, since we have represented with the red line only the $\bar{\alpha}$ and $\bar{\beta}$ values for the unaltered LVG catalog. We do not explicitly show this here, but for each of the mock catalogues, HINORA finds again the \textit{same} ring as for the original data shown in Fig.~\ref{fig:map}, but with marginally different values that eventually give rise to the shown error bars. This alone is already a confirmation of the stability of the identified structure, something to be investigated in more quantitative detail now.\\

\begin{figure}
	\includegraphics[width=\columnwidth]{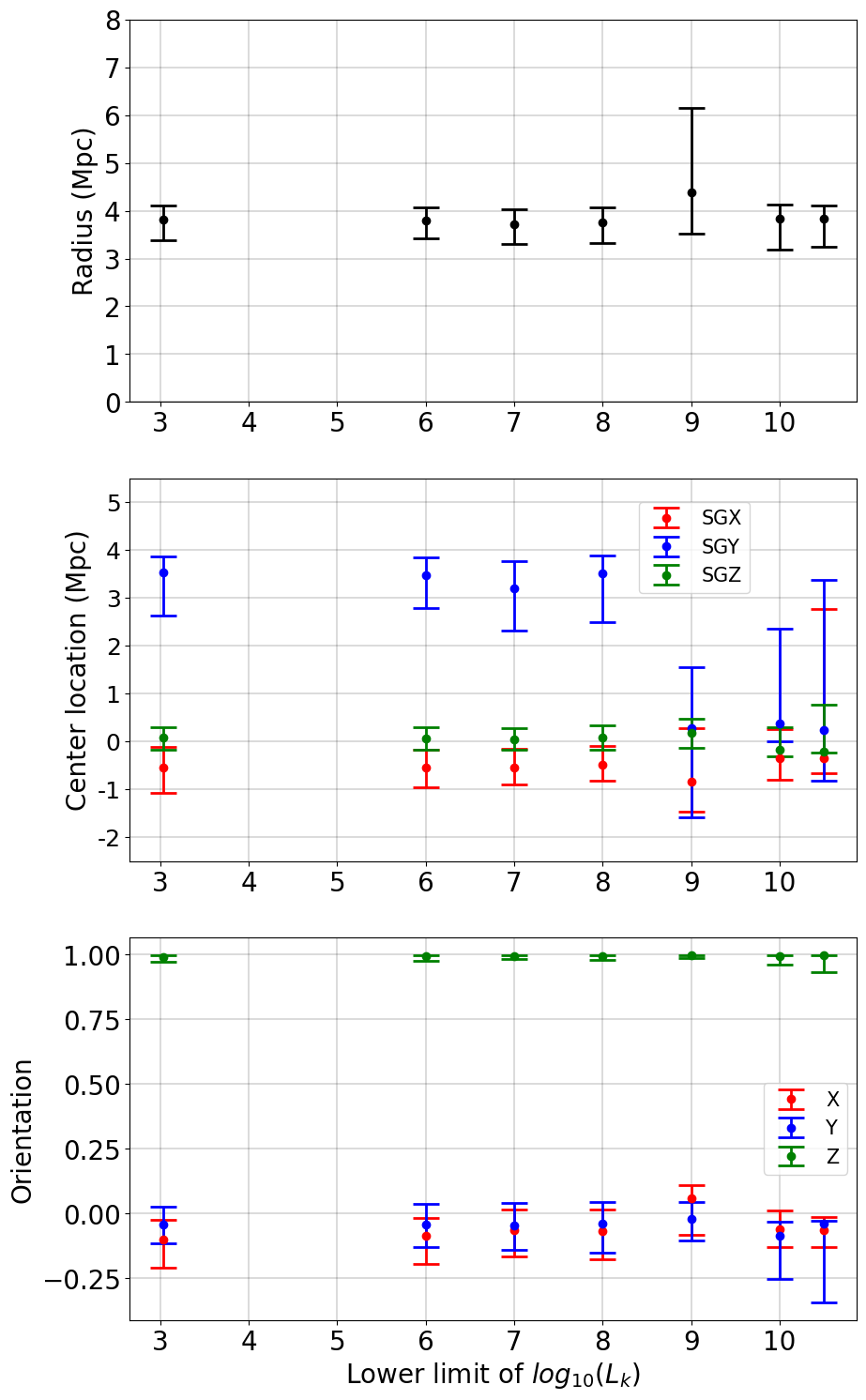} 
    \caption{Characteristics of the ring found by HINORA as a function of K-band cut. The upper panel shows the radius of the ring in Mpc, the middle panel the three components of the position vector of the ring centre in super-galactic coordinates, and the lower panel the three components of the unit vector normal to the plane defined by the ring.}
    \label{fig:modelsparam}
\end{figure}

Figure \ref{fig:modelsparam} shows the variation of ring radius (upper panel), ring centre (middle panel), and ring orientation (lower panel) with applied luminosity cut for $L_K$. The units are Mpc and given within the super-galactic frame, and for centre and orientation the three components of the respective vector are shown, which is further normalized to unity in the latter case of the vector normal to the plane defined by the ring. We note that the ring based upon the satellite galaxies (i.e. up to luminosity cut $\log_{10}L_K=8$) is remarkably stable. There are hardly any noticeable variations, and if there are they are certainly within the error bars. We further find a change for the cut $\log_{10}L_K=9$ as we have seen that this corresponds to another ring, a one akin to the original Council of Giants. But also for the Council ring we confirm stability within the error bars.\\

\begin{figure}
	\includegraphics[width=\columnwidth]{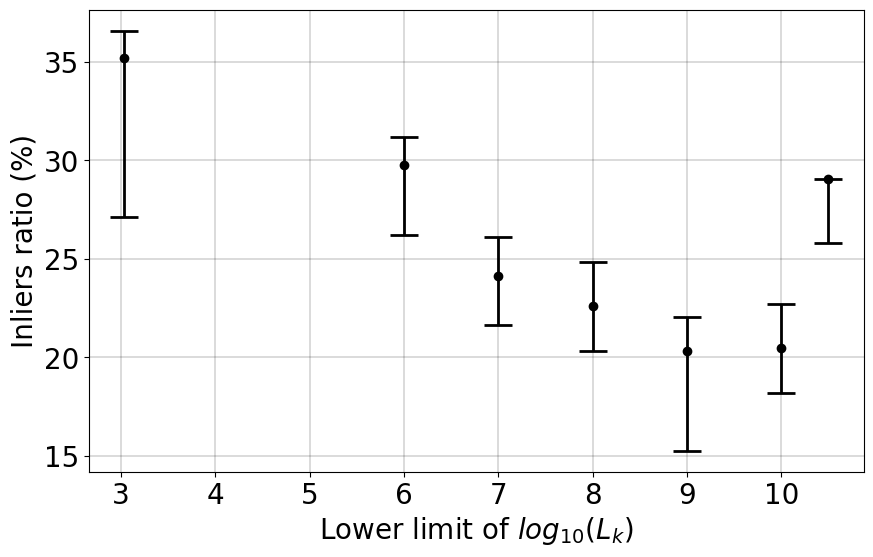} 
    \caption{Percentage of data belonging to the model (inliers) found by HINORA for different LVG luminosity cutoffs. The values are the mean of the method repetition, and the error corresponds to $85\%$ of the cases.}
    \label{fig:percent}
\end{figure}

We have seen (qualitatively) in Fig.~\ref{fig:map} that the number of galaxies in the identified ring reduces when increasing the luminosity cut. We now like to quantify this by showing in Fig.~\ref{fig:percent} the percentage of inliers as a function of $\log_{10}L_K$. In order to verify the actual number of ring member galaxies, one would need to compare this against all the galaxies in the sample for a given cut as presented in Fig.~\ref{fig:luminosity}. It is obvioues that, as $\log_{10}(L_{K})$ increases, the number of ring galaxies decreases. We also note that for the last two cuts that focus on the Council of Giants rings the inlier ratio rises again. This is due to the fact that there are only very few such bright galaxies in the whole sample, with of order 25 - 30 per cent making up the ring. But it is also remarkable that ca. 30 per cent of all galaxies from the LVG catalogue (within 10 Mpc distance to the MW, no cut applied) are forming part of the `satellite ring'. Such a structure should manifest itself not only in our ring-finding approach, but should in fact also emerge in other quantifications of possible structures, such as -- for instance -- the two-point correlation function (2PCF), as already reported by \citet{Tully1978}: even though not directly related to the structure finding algorithm HINORA, but motivated by the apparent existence of two distinct ring-like features in the Local Universe, we also calculated the 2PCF\footnote{For the calculation of the 2PCF we apply the code provided by \citet[][\url{https://www.astroml.org/user_guide/correlation_functions.html}]{astroML}.} for the same subset of LVG galaxies as used in the previous figures. The result is explained and shown in \ref{2PCF}.

\section{Discussion of HINORA}
\label{sec:discussion}
Throughout this work, we have described a computational method that allows us to identify simple patterns in complex point clouds, where most of the data may be noise. By analyzing the density and dispersion of data, this algorithm not only provides us with a way to locate specific distributions in arbitrarily large databases, but also gives us the intrinsic characteristics of the patterns we are looking for. By applying it to search for CoG-like galaxy rings, we detected two such structures that maintain their properties stable in their respective mass regimes. The `satellite ring' was found to be stable in the K-band luminosity range $log_{10}(L_{K}) > [3,...,9]$ cuts, and the CoG predominates for the more massive galaxies between $log_{10}(L_{K}) > 9$ and $10.5$). While the `satellite ring' shows higher stability and smaller variations in its characteristics over the wide mass range in which it dominates, the `council' nevertheless has better statistical properties such as higher uniformity and lower noise. The differences between the two rings can be explained by the influence of the haloes on the distribution of matter they contain, because smaller galaxies tend to cluster around certain massive galaxies that form specific groups. 

Before taking the next step and applying to HINORA to other, larger datasets, it is essential to establish the capabilities and limitations of the method used in order to correctly interpret the results it provides. The operating range of HINORA has been tested in the \ref{tests}, which shows that this method loses efficiency below $8\%$ inliers for a density of 100 points in a spherical volume of radius 10 in arbitrary units. The lowest ratios achieved in our analysis are found at the $log_{10}(L_{K})> 9$ cutoff, with 118 galaxies and approximately $15\%$ inliers. This ensures a high level of reliability close to $100\%$ in the regime in which we have used HINORA so far, i.e. in the LVG catalogue. Note that the method has high resilience to noise due to the fact that in its operating regime it lacks false positives, thus we will have the same level of efficiency if we double the analysis volume but without adding more inliers (thus achieving an accuracy of $4\%$). Since this exercise can be performed an arbitrary number of times without losing efficiency, the HINORA algorithm is able to deal with arbitrarily large noise, at the cost of a logarithmic increase in the number of iterations governed by equation \ref{ec:methodec1}. Given the generalizability, the method is potentially useful for any field that requires locating simple patterns in large databases where most of them are not of interest.

One of the reasons why HINORA may be defective for low densities (or few inliers) is its need for the existence of certain points forming the desired figure. For this reason, for ranges that only take into account the most luminous galaxies, the method loses efficiency. This is because decreasing the density of points decreases the probability that a subgroup of 3 points will accurately generate the model. In the case of CoG, it is unfeasible to $100\%$ recreate it. However, the method does not seek to be precise but resilient to noise. In this sense, the results for the last cuts lose precision given the performance of HINORA itself, although this is negligible compared to the uncertainty in the estimated distance for the galaxies.

Another important factor to consider is the possibility that more than one ring may be considered correct for the same cut-off in luminosity. In Figs.~\ref{fig:modelsalphabeta} and \ref{fig:modelsparam} the main source of the error bars for the cuts smaller than $log_{10}(L_{K}) > 9$ are mainly coming from the error in the distance/luminosity estimated in LVG. However, for the cuts $log_{10}(L_{K})>10$ and $10.5$ the error bars exhibit noticeable anomalies in the lower left panel of Fig.\ref{fig:modelsparam}. The increase in the observed uncertainty is due to the fact that the models considered correct differ more in their location, to the point that they may even be formed by different galaxies. For the last two cuts, the shifts of the center toward higher SGY values suggest that the `satellite ring' is still weakly favored in these ranges.

We like to close this discussion with reminding the reader that HINORA is not limited to ring-like structures. But when searching for different patterns, one would need to bear in mind that this also entails changes to our quality assessment that currently assume spherical symmetry. But in that regards, $\alpha$ is fully generalizable since one would only have to change the shape of the volume of the environment depending on the figure to be studied. For example, if we are looking for planes, the environment to be studied is a box. Further, $\beta$ uses generalized coordinates (since it is a normalized and dimensionless term). In our case we have used angles, but any other choice can be applied. For example, for straight lines we could use the distances that separate the projections of the points on the line; for a plane, the centroid strategy used in $\bar{\beta}$ can be employed.

\section{Summary \& Conclusions}
\label{sec:conclusions}
We have presented a novel method to detect patterns in 3D point distributions, called HINORA (HIgh-NOise RANdom SAmple Consensus). For the moment we have restricted ourselves to search for ring-like structures, but this is easily extendable to any other geometrical patterns. The method makes little assumptions, but the mathematical description of the pattern to search for, and the size of the model, i.e. in our case the width of the 3D torus encompassing the ring-like feature. Our first target for applying HINORA to a set of science data was the Local Volume Galaxies catalogue \citep{Karachentsev2013}, the most complete catalogue of local galaxies that has been compiled to date. This choice was motivated by the observation that massive galaxies in the Local Universe arrange themselves as a so-called `Council of Giants' (CoG) \citep[][]{McCall2014}.

We confirm the existence of the Council of Giants, though -- according to our findings -- Maffei 1 and 2 apparently do not belong to it. This can be attributed to the different distance of these two galaxies in the LVG data compared to the data used by \citet{McCall2014}. However, HINORA has also found another ring-like structure that is primarily defined via low-mass satellite galaxies and hence does not have the isotropic distribution as the former CoG. Both these rings do lie within the Local Sheet and it also needs to be said that these are the only rings detected by HINORA. It remains to be seen which of these two rings (i.e. the `satellite ring' or the `Council') do in fact relate to structures of astrophysical interest. They might be mere chance configurations or actually reflecting something more profound in the formation history of the Local Universe. For the moment the detection of them in the LVG data should be interpreted with care. Their existence rather verifies the mode of operation of HINORA, and it yet remains to be seen whether this is of physical significance or not.


As the application of the novel structure finder HINORA to astrophysical data left us with more open questions than answers, we aim applying HINORA also to simulations of cosmic structure formation. First and foremost we will use the constrained Local Universe simulations as provided by the CLUES-HESTIA project \citep{Libeskind2020}. That suite of simulations resemble with great accuracy the actual observed Local Universe, including the correct properties for the Milky Way and M31. However, to fully understand the requirements to form such galaxy rings, we will also use `Local Group candidates' as found in the Illustris-TNG simulation \citep{Vogelsberger2014} as a control sample. This will allow for a deeper understanding of the physical implications of rings and whether they are indeed a common structure in the universe or a mere coincidental formation.

\begin{acknowledgement}
All authors thank the referee for their constructive comments that helped to improve the paper. EO thanks Ingridi Teixeira for her translation assistance and Clara Rebanal for her advice. AK thanks Biohazard for urban discipline. 
\end{acknowledgement}

\paragraph{Funding Statement}

EO and AK are supported by the Ministerio de Ciencia e Innovaci\'{o}n (MICINN) under research grant PID2021-122603NB-C21. DIM is supported by the grant \textnumero~075--15--2022--262 (13.MNPMU.21.0003) of the Ministry of Science and Higher Education of the Russian Federation.



\paragraph{Data Availability Statement}

The LVG data used here is publicly available from \url{https://www.sao.ru/lv/lvgdb/introduction.php}. The analysis tools (incl. the HINORA code) are provided upon reasonable request.

\printendnotes

\printbibliography 

@ARTICLE{Tully2023_BAO,
       author = {{Tully}, R. Brent and {Howlett}, Cullan and {Pomar{\`e}de}, Daniel},
        title = "{Ho'oleilana: An Individual Baryon Acoustic Oscillation?}",
      journal = {The Astrophysical Journal},
         year = 2023,
        month = sep,
       volume = {954},
       number = {2},
          eid = {169},
        pages = {169},
          doi = {10.3847/1538-4357/aceaf3},
archivePrefix = {arXiv},
       eprint = {2309.00677},
 primaryClass = {astro-ph.CO},
       adsurl = {https://ui.adsabs.harvard.edu/abs/2023ApJ...954..169T}
}

@ARTICLE{Frenk1983,
       author = {{Frenk}, C.~S. and {White}, S.~D.~M. and {Davis}, M.},
        title = "{Nonlinear evolution of large-scale structure in the universe}",
      journal = {The Astrophysical Journal},
         year = 1983,
        month = aug,
       volume = {271},
        pages = {417-430},
          doi = {10.1086/161209},
       adsurl = {https://ui.adsabs.harvard.edu/abs/1983ApJ...271..417F}
}

@ARTICLE{Davis1985,
       author = {{Davis}, M. and {Efstathiou}, G. and {Frenk}, C.~S. and {White}, S.~D.~M.},
        title = "{The evolution of large-scale structure in a universe dominated by cold dark matter}",
      journal = {apj},
         year = 1985,
        month = may,
       volume = {292},
        pages = {371-394},
          doi = {10.1086/163168},
       adsurl = {https://ui.adsabs.harvard.edu/abs/1985ApJ...292..371D}
}

@ARTICLE{Turner1975,
       author = {{Turner}, E.~L. and {Gott}, J.~R., III},
        title = "{Evidence for a spatially homogeneous component of the universe: single galaxies.}",
      journal = {The Astrophysical Journal Letters},
         year = 1975,
        month = may,
       volume = {197},
        pages = {L89-L93},
          doi = {10.1086/181785},
       adsurl = {https://ui.adsabs.harvard.edu/abs/1975ApJ...197L..89T}
}

@ARTICLE{Soneira1977,
       author = {{Soneira}, R.~M. and {Peebles}, P.~J.~E.},
        title = "{Is there evidence for a spatially homogeneous population of field galaxies}",
      journal = {The Astrophysical Journal},
         year = 1977,
        month = jan,
       volume = {211},
        pages = {1-15},
          doi = {10.1086/154898},
       adsurl = {https://ui.adsabs.harvard.edu/abs/1977ApJ...211....1S}
}

@ARTICLE{Libeskind2020,
       author = {{Libeskind}, Noam I. and {Carlesi}, Edoardo and {Grand}, Robert J.~J. and {Khalatyan}, Arman and {Knebe}, Alexander and {Pakmor}, Ruediger and {Pilipenko}, Sergey and {Pawlowski}, Marcel S. and {Sparre}, Martin and {Tempel}, Elmo and {Wang}, Peng and {Courtois}, H{\'e}l{\`e}ne M. and {Gottl{\"o}ber}, Stefan and {Hoffman}, Yehuda and {Minchev}, Ivan and {Pfrommer}, Christoph and {Sorce}, Jenny G. and {Springel}, Volker and {Steinmetz}, Matthias and {Tully}, R. Brent and {Vogelsberger}, Mark and {Yepes}, Gustavo},
        title = "{The HESTIA project: simulations of the Local Group}",
      journal = {mnras},
         year = 2020,
        month = oct,
       volume = {498},
       number = {2},
        pages = {2968-2983},
          doi = {10.1093/mnras/staa2541},
archivePrefix = {arXiv},
       eprint = {2008.04926},
 primaryClass = {astro-ph.GA},
       adsurl = {https://ui.adsabs.harvard.edu/abs/2020MNRAS.498.2968L}
}

@ARTICLE{Vogelsberger2014,
       author = {{Vogelsberger}, M. and {Genel}, S. and {Springel}, V. and {Torrey}, P. and {Sijacki}, D. and {Xu}, D. and {Snyder}, G. and {Bird}, S. and {Nelson}, D. and {Hernquist}, L.},
        title = "{Properties of galaxies reproduced by a hydrodynamic simulation}",
      journal = {nat},
         year = 2014,
        month = may,
       volume = {509},
       number = {7499},
        pages = {177-182},
          doi = {10.1038/nature13316},
archivePrefix = {arXiv},
       eprint = {1405.1418},
 primaryClass = {astro-ph.CO},
       adsurl = {https://ui.adsabs.harvard.edu/abs/2014Natur.509..177V}
}

@ARTICLE{Holtzman1989,
       author = {{Holtzman}, Jon A.},
        title = "{Microwave Background Anisotropies and Large-Scale Structure in Universes with Cold Dark Matter, Baryons, Radiation, and Massive and Massless Neutrinos}",
      journal = {The Astrophysical Journal Supplement Series},
         year = 1989,
        month = sep,
       volume = {71},
        pages = {1},
          doi = {10.1086/191362},
       adsurl = {https://ui.adsabs.harvard.edu/abs/1989ApJS...71....1H}
}

@ARTICLE{Bond1984,
       author = {{Bond}, J.~R. and {Efstathiou}, G.},
        title = "{Cosmic background radiation anisotropies in universes dominated by nonbaryonic dark matter}",
      journal = {The Astrophysical Journal Letters},
         year = 1984,
        month = oct,
       volume = {285},
        pages = {L45-L48},
          doi = {10.1086/184362},
       adsurl = {https://ui.adsabs.harvard.edu/abs/1984ApJ...285L..45B}
}

@ARTICLE{Peebles1970,
       author = {{Peebles}, P.~J.~E. and {Yu}, J.~T.},
        title = "{Primeval Adiabatic Perturbation in an Expanding Universe}",
      journal = {apj},
         year = 1970,
        month = dec,
       volume = {162},
        pages = {815},
          doi = {10.1086/150713},
       adsurl = {https://ui.adsabs.harvard.edu/abs/1970ApJ...162..815P}
}

@ARTICLE{Huchra1983,
       author = {{Huchra}, J. and {Davis}, M. and {Latham}, D. and {Tonry}, J.},
        title = "{A survey of galaxy redshifts. IV - The data}",
      journal = {The Astrophysical Journal Supplement Series},
         year = 1983,
        month = jun,
       volume = {52},
        pages = {89-119},
          doi = {10.1086/190860},
       adsurl = {https://ui.adsabs.harvard.edu/abs/1983ApJS...52...89H}
}

@ARTICLE{Ziparo2016,
       author = {{Ziparo}, F. and {Smith}, G.~P. and {Mulroy}, S.~L. and {Lieu}, M. and {Willis}, J.~P. and {Hudelot}, P. and {McGee}, S.~L. and {Fotopoulou}, S. and {Lidman}, C. and {Lavoie}, S. and {Pierre}, M. and {Adami}, C. and {Chiappetti}, L. and {Clerc}, N. and {Giles}, P. and {Maughan}, B. and {Pacaud}, F. and {Sadibekova}, T.},
        title = "{The XXL Survey. X. K-band luminosity - weak-lensing mass relation for groups and clusters of galaxies}",
      journal = {Astronomy \& Astrophysics},
         year = 2016,
        month = jun,
       volume = {592},
          eid = {A9},
        pages = {A9},
          doi = {10.1051/0004-6361/201526792},
archivePrefix = {arXiv},
       eprint = {1512.03903},
 primaryClass = {astro-ph.GA},
       adsurl = {https://ui.adsabs.harvard.edu/abs/2016A&A...592A...9Z}
}

@ARTICLE{Jarrett2013,
       author = {{Jarrett}, T.~H. and {Masci}, F. and {Tsai}, C.~W. and {Petty}, S. and {Cluver}, M.~E. and {Assef}, Roberto J. and {Benford}, D. and {Blain}, A. and {Bridge}, C. and {Donoso}, E. and {Eisenhardt}, P. and {Koribalski}, B. and {Lake}, S. and {Neill}, James D. and {Seibert}, M. and {Sheth}, K. and {Stanford}, S. and {Wright}, E.},
        title = "{Extending the Nearby Galaxy Heritage with WISE: First Results from the WISE Enhanced Resolution Galaxy Atlas}",
      journal = {aj},
         year = 2013,
        month = jan,
       volume = {145},
       number = {1},
          eid = {6},
        pages = {6},
          doi = {10.1088/0004-6256/145/1/6},
archivePrefix = {arXiv},
       eprint = {1210.3628},
 primaryClass = {astro-ph.CO},
       adsurl = {https://ui.adsabs.harvard.edu/abs/2013AJ....145....6J}
}

@ARTICLE{Zeldovich1970,
       author = {{Zel'dovich}, Ya. B.},
        title = "{Gravitational instability: An approximate theory for large density perturbations.}",
      journal = {Astronomy \& Astrophysics},
         year = 1970,
        month = mar,
       volume = {5},
        pages = {84-89},
       adsurl = {https://ui.adsabs.harvard.edu/abs/1970A&A.....5...84Z}
}

@article{Guth1981,
  title = {Inflationary universe: A possible solution to the horizon and flatness problems},
  author = {Guth, Alan H.},
  journal = {Phys. Rev. D},
  volume = {23},
  issue = {2},
  pages = {347--356},
  numpages = {0},
  year = {1981},
  month = {Jan},
  publisher = {American Physical Society},
  doi = {10.1103/PhysRevD.23.347},
  url = {https://link.aps.org/doi/10.1103/PhysRevD.23.347}
}

@article{Linde1982,
title = {A new inflationary universe scenario: A possible solution of the horizon, flatness, homogeneity, isotropy and primordial monopole problems},
journal = {Physics Letters B},
volume = {108},
number = {6},
pages = {389-393},
year = {1982},
issn = {0370-2693},
doi = {https://doi.org/10.1016/0370-2693(82)91219-9},
url = {https://www.sciencedirect.com/science/article/pii/0370269382912199},
author = {A.D. Linde}
}

@article{Lapparent1986,
  title={A slice of the universe},
  author={De Lapparent, Val{\'e}rie and Geller, Margaret J and Huchra, John P},
  journal={Astrophysical Journal, Part 2-Letters to the Editor (ISSN 0004-637X), vol. 302, March 1, 1986, p. L1-L5. Research supported by the Smithsonian Institution.},
  volume={302},
  pages={L1--L5},
  year={1986}
}

@article{Eisenstein2005,
doi = {10.1086/466512},
url = {https://dx.doi.org/10.1086/466512},
year = {2005},
month = {nov},
publisher = {},
volume = {633},
number = {2},
pages = {560},
author = {Daniel J. Eisenstein and Idit Zehavi and David W. Hogg and Roman Scoccimarro and Michael R. Blanton and Robert C. Nichol and Ryan Scranton and Hee-Jong Seo and Max Tegmark and Zheng Zheng and Scott F. Anderson and Jim Annis and Neta Bahcall and Jon Brinkmann and Scott Burles and Francisco J. Castander and Andrew Connolly and Istvan Csabai and Mamoru Doi and Masataka Fukugita and Joshua A. Frieman and Karl Glazebrook and James E. Gunn and John S. Hendry and Gregory Hennessy and Zeljko Ivezić and Stephen Kent and Gillian R. Knapp and Huan Lin and Yeong-Shang Loh and Robert H. Lupton and Bruce Margon and Timothy A. McKay and Avery Meiksin and Jeffery A. Munn and Adrian Pope and Michael W. Richmond and David Schlegel and Donald P. Schneider and Kazuhiro Shimasaku and Christopher Stoughton and Michael A. Strauss and Mark SubbaRao and Alexander S. Szalay and István Szapudi and Douglas L. Tucker and Brian Yanny and Donald G. York},
title = {Detection of the Baryon Acoustic Peak in the Large-Scale Correlation Function of SDSS Luminous Red Galaxies},
journal = {The Astrophysical Journal}
}

@INPROCEEDINGS{Tully1978,
       author = {{Tully}, R.~B. and {Fisher}, J.~R.},
        title = "{Nearby Small Groups of Galaxies}",
    booktitle = {Large Scale Structures in the Universe},
       series = {Proceedings of the IAU Symposium},
         year = 1978,
       editor = {{Longair}, M.~S. and {Einasto}, J.},
       volume = {79},
        month = jan,
        pages = {31},
       adsurl = {https://ui.adsabs.harvard.edu/abs/1978IAUS...79...31T}
}

@article{Karachentsev2018,
	doi = {10.1002/asna.201813520},
  
	url = {https://doi.org/10.1002{\%}2Fasna.201813520},
  
	year = 2018,
	month = {aug},
  
	publisher = {Wiley},
  
	volume = {339},
  
	number = {7-8},
  
	pages = {615--622},
  
	author = {I. D. Karachentsev and K. N. Telikova},
  
	title = {Stellar and dark matter density in the Local Universe},
  
	journal = {Astronomische Nachrichten}
}

@article{Tully2008,
doi = {10.1086/527428},
url = {https://dx.doi.org/10.1086/527428},
year = {2008},
month = {mar},
publisher = {},
volume = {676},
number = {1},
pages = {184},
author = {R. Brent Tully and Edward J. Shaya and Igor D. Karachentsev and Hélène M. Courtois and Dale D. Kocevski and Luca Rizzi and Alan Peel},
title = {Our Peculiar Motion Away from the Local Void},
journal = {The Astrophysical Journal}
}

@inproceedings{Choi2009,
author = {Choi, Sunglok and Kim, Taemin and Yu, Wonpil},
year = {2009},
month = {01},
pages = {},
title = {Performance evaluation of RANSAC family},
volume = {24},
journal = {Proceedings of the British Machine Vision Conference 2009},
doi = {10.5244/C.23.81}
}

@article{Karachentsev2013,
doi = {10.1088/0004-6256/145/4/101},
url = {https://dx.doi.org/10.1088/0004-6256/145/4/101},
year = {2013},
month = {mar},
publisher = {The American Astronomical Society},
volume = {145},
number = {4},
pages = {101},
author = {Igor D. Karachentsev and Dmitry I. Makarov and Elena I. Kaisina},
title = {UPDATED NEARBY GALAXY CATALOG},
journal = {The Astronomical Journal}
}

@article{Neuzil2020,
    author = {Neuzil, Maria K and Mansfield, Philip and Kravtsov, Andrey V},
    title = "{The Sheet of Giants: Unusual properties of the Milky Way’s immediate neighbourhood}",
    journal = {Monthly Notices of the Royal Astronomical Society},
    volume = {494},
    number = {2},
    pages = {2600-2617},
    year = {2020},
    month = {04},
    issn = {0035-8711},
    doi = {10.1093/mnras/staa898},
    url = {https://doi.org/10.1093/mnras/staa898},
    eprint = {https://academic.oup.com/mnras/article-pdf/494/2/2600/33113646/staa898.pdf},
}

@article{McCall2014,
    author = {McCall, Marshall L.},
    title = "{A Council of Giants}",
    journal = {Monthly Notices of the Royal Astronomical Society},
    volume = {440},
    number = {1},
    pages = {405-426},
    year = {2014},
    month = {02},
    issn = {0035-8711},
    doi = {10.1093/mnras/stu199},
    url = {https://doi.org/10.1093/mnras/stu199},
    eprint = {https://academic.oup.com/mnras/article-pdf/440/1/405/9376493/stu199.pdf},
}

@article{Jarrett2000,
doi = {10.1086/301330},
url = {https://dx.doi.org/10.1086/301330},
year = {2000},
month = {may},
publisher = {},
volume = {119},
number = {5},
pages = {2498},
author = {T. H. Jarrett and T. Chester and R. Cutri and S. Schneider and M. Skrutskie and J. P. Huchra},
title = {2MASS Extended Source Catalog: Overview and Algorithms},
journal = {The Astronomical Journal}
}

@article{Jarrett2003,
doi = {10.1086/345794},
url = {https://dx.doi.org/10.1086/345794},
year = {2003},
month = {feb},
publisher = {},
volume = {125},
number = {2},
pages = {525},
author = {T. H. Jarrett and T. Chester and R. Cutri and S. E. Schneider and J. P. Huchra},
title = {The 2MASS Large Galaxy Atlas},
journal = {The Astronomical Journal}
}

@article{Fingerhut2010,
doi = {10.1088/0004-637X/716/1/792},
url = {https://dx.doi.org/10.1088/0004-637X/716/1/792},
year = {2010},
month = {may},
publisher = {The American Astronomical Society},
volume = {716},
number = {1},
pages = {792},
author = {Robin L. Fingerhut and Marshall L. McCall and Mauricio Argote and Michelle E. Cluver and Shogo Nishiyama and Rami T. F. Rekola and Michael G. Richer and Ovidiu Vaduvescu and Patrick A. Woudt},
title = {DEEP Ks-NEAR-INFRARED SURFACE PHOTOMETRY OF 80 DWARF IRREGULAR GALAXIES IN THE LOCAL VOLUME},
journal = {The Astrophysical Journal}
}

@article{Vaduvescu2005,
doi = {10.1086/444498},
url = {https://dx.doi.org/10.1086/444498},
year = {2005},
month = {oct},
publisher = {},
volume = {130},
number = {4},
pages = {1593},
author = {Ovidiu Vaduvescu and Marshall L. McCall and Michael G. Richer and Robin L. Fingerhut},
title = {Infrared Properties of Star-forming Dwarf Galaxies. I. Dwarf Irregular Galaxies in the Local Volume*},
journal = {The Astronomical Journal}
}

@article{Vaduvescu2006,
doi = {10.1086/498723},
url = {https://dx.doi.org/10.1086/498723},
year = {2006},
month = {mar},
publisher = {},
volume = {131},
number = {3},
pages = {1318},
author = {Ovidiu Vaduvescu and Michael G. Richer and Marshall L. McCall},
title = {Infrared Properties of Star-forming Dwarf Galaxies. II. Blue Compact Dwarf Galaxies in the Virgo Cluster*},
journal = {The Astronomical Journal}
}

@article{Bell2001,
doi = {10.1086/319728},
url = {https://dx.doi.org/10.1086/319728},
year = {2001},
month = {mar},
publisher = {},
volume = {550},
number = {1},
pages = {212},
author = {Eric F. Bell and Roelof S. de Jong},
title = {Stellar Mass-to-Light Ratios and the
Tully-Fisher Relation},
journal = {The Astrophysical Journal}
}

@article{Bell2003,
doi = {10.1086/378847},
url = {https://dx.doi.org/10.1086/378847},
year = {2003},
month = {dec},
publisher = {},
volume = {149},
number = {2},
pages = {289},
author = {Eric F. Bell and Daniel H. McIntosh and Neal Katz and Martin D. Weinberg},
title = {The Optical and Near-Infrared Properties of Galaxies. I.
Luminosity and Stellar Mass Functions},
journal = {The Astrophysical Journal Supplement Series}
}

@article{Beare2019,
doi = {10.3847/1538-4357/ab041a},
url = {https://dx.doi.org/10.3847/1538-4357/ab041a},
year = {2019},
month = {mar},
publisher = {The American Astronomical Society},
volume = {873},
number = {1},
pages = {78},
author = {Richard Beare and Michael J. I. Brown and Kevin Pimbblet and Edward N. Taylor},
title = {Evolution of the Stellar Mass Function and Infrared Luminosity Function of Galaxies since z=1.2},
journal = {The Astrophysical Journal}
}

@article{Kirby2008,
doi = {10.1088/0004-6256/136/5/1866},
url = {https://dx.doi.org/10.1088/0004-6256/136/5/1866},
year = {2008},
month = {oct},
publisher = {The American Astronomical Society},
volume = {136},
number = {5},
pages = {1866},
author = {Emma M. Kirby and Helmut Jerjen and Stuart D. Ryder and Simon P. Driver},
title = {DEEP NEAR-INFRARED SURFACE PHOTOMETRY OF 57 GALAXIES IN THE LOCAL SPHERE OF INFLUENCE},
journal = {The Astronomical Journal}
}

@article{Fischler1981,
author = {Fischler, Martin A. and Bolles, Robert C.},
title = {Random Sample Consensus: A Paradigm for Model Fitting with Applications to Image Analysis and Automated Cartography},
year = {1981},
issue_date = {June 1981},
publisher = {Association for Computing Machinery},
address = {New York, NY, USA},
volume = {24},
number = {6},
issn = {0001-0782},
url = {https://doi.org/10.1145/358669.358692},
doi = {10.1145/358669.358692},
journal = {Commun. ACM},
month = {jun},
pages = {381–395},
numpages = {15}
}

@InProceedings{Raguram2008,
author="Raguram, Rahul
and Frahm, Jan-Michael
and Pollefeys, Marc",
editor="Forsyth, David
and Torr, Philip
and Zisserman, Andrew",
title="A Comparative Analysis of RANSAC Techniques Leading to Adaptive Real-Time Random Sample Consensus",
booktitle="Computer Vision -- ECCV 2008",
year="2008",
publisher="Springer Berlin Heidelberg",
address="Berlin, Heidelberg",
pages="500--513",
isbn="978-3-540-88688-4"
}

@article{Raguram2013,
  title={USAC: A Universal Framework for Random Sample Consensus},
  author={Rahul Raguram and Ondřej Chum and Marc Pollefeys and Jiri Matas and Jan-Michael Frahm},
  journal={IEEE Transactions on Pattern Analysis and Machine Intelligence},
  year={2013},
  volume={35},
  pages={2022-2038}
}

@article{Xie2019,
title={Linking Points With Labels in 3D: A Review of
Point Cloud Semantic Segmentation},
author={Xie, Yuxing and Tian, Jiaojiao and Zhu, Xiao
Xiang},
journal={IEEE Geoscience and Remote Sensing Magazine},
year={2020},
publisher={IEEE},
doi={10.1109/MGRS.2019.2937630}
}

@article{Matas2004,
author = {Matas, Jiri and Chum, O},
year = {2004},
month = {09},
pages = {837-842},
title = {Randomized RANSAC with Td,d test},
volume = {22},
journal = {Image and Vision Computing},
doi = {10.1016/j.imavis.2004.02.009}
}

@article{Anand2019,
doi = {10.3847/2041-8213/aafee6},
url = {https://dx.doi.org/10.3847/2041-8213/aafee6},
year = {2019},
month = {feb},
publisher = {The American Astronomical Society},
volume = {872},
number = {1},
pages = {L4},
author = {Gagandeep S. Anand and R. Brent Tully and Luca Rizzi and Igor D. Karachentsev},
title = {The Distance and Motion of the Maffei Group},
journal = {The Astrophysical Journal Letters}
}

@INPROCEEDINGS{astroML,
  author={{Vanderplas}, J.T. and {Connolly}, A.J.
          and {Ivezi{\'c}}, {\v Z}. and {Gray}, A.},
  booktitle={Conference on Intelligent Data Understanding (CIDU)},
  title={Introduction to astroML: Machine learning for astrophysics},
  month={oct.},
  pages={47 -54},
  doi={10.1109/CIDU.2012.6382200},
  year={2012}
}

@article{Tully2016,
doi = {10.3847/0004-6256/152/2/50},
url = {https://dx.doi.org/10.3847/0004-6256/152/2/50},
year = {2016},
month = {aug},
publisher = {The American Astronomical Society},
volume = {152},
number = {2},
pages = {50},
author = {R. Brent Tully and Hélène M. Courtois and Jenny G. Sorce},
title = {COSMICFLOWS-3},
journal = {The Astronomical Journal}
}

@article{
Geller1989,
author = {Margaret J. Geller  and John P. Huchra },
title = {Mapping the Universe},
journal = {Science},
volume = {246},
number = {4932},
pages = {897-903},
year = {1989},
doi = {10.1126/science.246.4932.897},
URL = {https://www.science.org/doi/abs/10.1126/science.246.4932.897},
eprint = {https://www.science.org/doi/pdf/10.1126/science.246.4932.897}}

@ARTICLE{2019AstBu..74..111K,
       author = {{Karachentsev}, I.~D. and {Kaisina}, E.~I.},
        title = "{Dwarf Galaxies in the Local Volume}",
      journal = {Astrophysical Bulletin},
         year = {2019},
        month = {apr},
       volume = {74},
       number = {2},
        pages = {111-127},
          doi = {10.1134/S1990341319020019},
archivePrefix = {arXiv},
       eprint = {1905.08477},
 primaryClass = {astro-ph.GA},
       adsurl = {https://ui.adsabs.harvard.edu/abs/2019AstBu..74..111K}
}

@article{Tully2023_CF4,
doi = {10.3847/1538-4357/ac94d8},
url = {https://dx.doi.org/10.3847/1538-4357/ac94d8},
year = {2023},
month = {feb},
publisher = {The American Astronomical Society},
volume = {944},
number = {1},
pages = {94},
author = {R. Brent Tully and Ehsan Kourkchi and Hélène M. Courtois and Gagandeep S. Anand and John P. Blakeslee and Dillon Brout and Thomas de Jaeger and Alexandra Dupuy and Daniel Guinet and Cullan Howlett and Joseph B. Jensen and Daniel Pomarède and Luca Rizzi and David Rubin and Khaled Said and Daniel Scolnic and Benjamin E. Stahl},
title = {Cosmicflows-4},
journal = {The Astrophysical Journal}
}

@article{Cole2005,
    author = {Cole, Shaun and Percival, Will J. and Peacock, John A. and Norberg, Peder and Baugh, Carlton M. and Frenk, Carlos S. and Baldry, Ivan and Bland-Hawthorn, Joss and Bridges, Terry and Cannon, Russell and Colless, Matthew and Collins, Chris and Couch, Warrick and Cross, Nicholas J. G. and Dalton, Gavin and Eke, Vincent R. and de Propris, Roberto and Driver, Simon P. and Efstathiou, George and Ellis, Richard S. and Glazebrook, Karl and Jackson, Carole and Jenkins, Adrian and Lahav, Ofer and Lewis, Ian and Lumsden, Stuart and Maddox, Steve and Madgwick, Darren and Peterson, Bruce A. and Sutherland, Will and Taylor, Keith and The 2dFGRS Team},
    title = "{The 2dF Galaxy Redshift Survey: power-spectrum analysis of the final data set and cosmological implications}",
    journal = {Monthly Notices of the Royal Astronomical Society},
    volume = {362},
    number = {2},
    pages = {505-534},
    year = {2005},
    month = {09},
    issn = {0035-8711},
    doi = {10.1111/j.1365-2966.2005.09318.x},
    url = {https://doi.org/10.1111/j.1365-2966.2005.09318.x},
    eprint = {https://academic.oup.com/mnras/article-pdf/362/2/505/6155670/362-2-505.pdf},
}

@article{Percival2001,
    author = {Percival, Will J. and Baugh, Carlton M. and Bland-Hawthorn, Joss and Bridges, Terry and Cannon, Russell and Cole, Shaun and Colless, Matthew and Collins, Chris and Couch, Warrick and Dalton, Gavin and De Propris, Roberto and Driver, Simon P. and Efstathiou, George and Ellis, Richard S. and Frenk, Carlos S. and Glazebrook, Karl and Jackson, Carole and Lahav, Ofer and Lewis, Ian and Lumsden, Stuart and Maddox, Steve and Moody, Stephen and Norberg, Peder and Peacock, John A. and Peterson, Bruce A. and Sutherland, Will and Taylor, Keith},
    title = "{The 2dF Galaxy Redshift Survey: the power spectrum and the matter content of the Universe}",
    journal = {Monthly Notices of the Royal Astronomical Society},
    volume = {327},
    number = {4},
    pages = {1297-1306},
    year = {2001},
    month = {11},
    issn = {0035-8711},
    doi = {10.1046/j.1365-8711.2001.04827.x},
    url = {https://doi.org/10.1046/j.1365-8711.2001.04827.x},
    eprint = {https://academic.oup.com/mnras/article-pdf/327/4/1297/3279954/327-4-1297.pdf},
}

@BOOK{Tully1987,
       author = {{Tully}, R. Brent and {Fisher}, J. Richard},
        title = "{Atlas of Nearby Galaxies}",
         year = 1987,
       adsurl = {https://ui.adsabs.harvard.edu/abs/1987ang..book.....T}
}

@ARTICLE{Tikhonov2018,
       author = {{Tikhonov}, N.~A. and {Galazutdinova}, O.~A.},
        title = "{Does the IC 342/Maffei Galaxy Group Really Exist?}",
      journal = {Astrophysical Bulletin},
         year = 2018,
        month = jul,
       volume = {73},
       number = {3},
        pages = {279-292},
          doi = {10.1134/S1990341318030021},
       adsurl = {https://ui.adsabs.harvard.edu/abs/2018AstBu..73..279T}
}

@ARTICLE{Kaisina2012,
       author = {{Kaisina}, E.~I. and {Makarov}, D.~I. and {Karachentsev}, I.~D. and {Kaisin}, S.~S.},
        title = "{Observational database for studies of nearby universe}",
      journal = {Astrophysical Bulletin},
         year = 2012,
        month = jan,
       volume = {67},
       number = {1},
        pages = {115-122},
          doi = {10.1134/S1990341312010105},
       adsurl = {https://ui.adsabs.harvard.edu/abs/2012AstBu..67..115K}
}

\appendix

\section{Operating range of HINORA}

\label{tests}
\begin{figure}
	\includegraphics[width=\columnwidth]{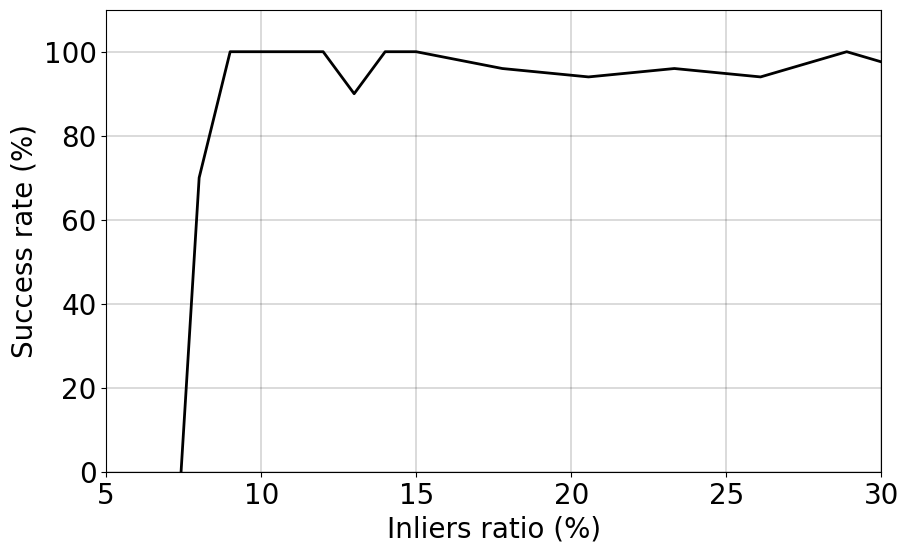} 
    \caption{Placing perfect rings into otherwise random data: HINORA's success rate for finding a ring-like structure as a function of percentage of inliers.}
    \label{fig:appendix1}
\end{figure}

To test the HINORA method described in Sec.~\ref{sec:method}, we have performed two tests to validate the range of its credibility. The HINORA settings are identical to the ones used in the main part of the study, but lowering the minimum number of inliers. The first test consists of determining the performance limit for a constant density of 100 points in a volume of 10 radius units, where the inliers ratio drops smoothly. For this purpose, we generate random data sets in which we introduce perfect circles (i.e. equidistant points on the circumference with constant radius) with randomly chosen orientation, location and radius. As we decrease the number of points composing this ring, we should see a limit where the success ratio of HINORA decreases when applied to find the rings. We define the success rate as the percentage of inliers found by HINORA with respects to the ones placed into the ring. A ring is considered successful when the ratio of inliers that HINORA finds is above 90 per cent. A failed ring has either a smaller inliers ratio is a test where a ring is passed or is not found at all (which is the most common case for failure). In Fig.~\ref{fig:appendix1} we now show this success rate as a function of the percentage of manually placed inliers with respects to the total number of (random) points in the test data. We observe that, if the ring contains less than ca. 8 per cent of the full data, HINORA is not able to find it anymore, even though it is a perfect ring. We are therefore confident that all the rings found and analysed in this work were correctly found as the inliers ratio for them is always above 10 per cent of the respective full data sample (see Fig.~\ref{fig:percent}).

\begin{figure}
	\includegraphics[width=\columnwidth]{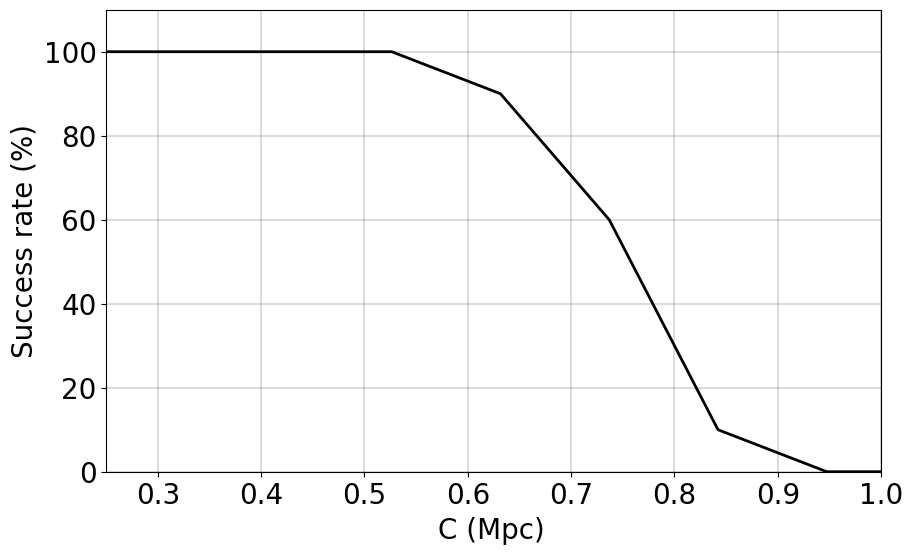} 
    \caption{Placing imperfect rings into otherwise random data: HINORA's success rate for finding a ring-like structure as a function of distance $C$ (in Mpc) to the perfect ring circumference.}
    \label{fig:appendix2}

\end{figure}

The second test consists of determining how the algorithm reacts to more realistic rings. To this extent we place rings into the same random data as used in the previous test, but now allowing for deviations by adding a random vector to each of the otherwise on a perfect circle placed points

\begin{equation}
    \vec{r}_i \rightarrow \vec{r}_i + C \ \vec{e} \ ,
\end{equation}
where $\vec{e}$ is a unit vector with random orientation, and $C$ measures the distance from the perfect ring circumference of radius 10 Mpc. For the test presented in Fig.~\ref{fig:appendix2} we continuously vary $C$ between 0 and 1. That figure clearly shows how the success rate to find a ring drops towards zero for $C\rightarrow1$, noting that we used the same $\tau=1$Mpc as in the main body of the paper, which then in turn tells us that rings will only be successfully found when the inliers are within a torus of $\pm0.5$Mpc. We further like to remark these tests were performed using a constant inliers ratio of 20 per cent of the data, and the algorithm uses $\bar{N_I}$ = 15 per cent of the data. 
Despite these tests, there are experimental gaps that produce alterations that are difficult to quantify. Observational effects like ZoA, involving apparent over/under density in the data, can alter the $\alpha$ value of the models near these conflict zones, and have to be studied individually when interpreting the results. In our case, no LVG ring is seriously affected by this type of gap caused by ZoA, although of importance are the distance errors in this region, which have been taken into account in the error bars in the figures of section \ref{sec:LVG}.

\section{Two-Point Correlation Analysis}
\label{2PCF}

\begin{figure}
	\includegraphics[width=\columnwidth]{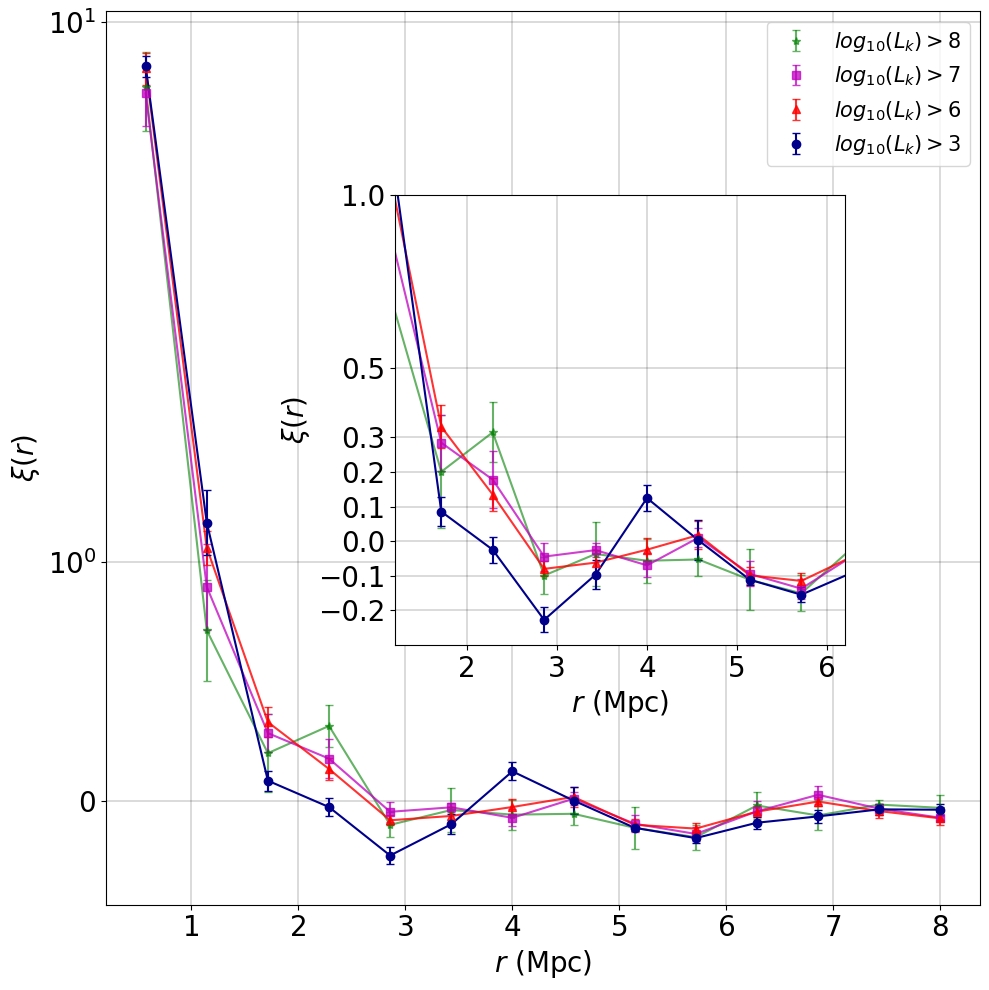} 
    \caption{Two-point correlation function applied to four K-band luminosity cuts. The inset panel is a zoom into the $r$-range $[1,6]$~Mpc.}
    \label{fig:2pcf}
\end{figure}

The significant presence of the rings detected by HINORA in the LVG data motivates to look for their footprint in statistical functions. In Fig.~\ref{fig:2pcf} we show the 2PCF for the luminosity cuts up to (and including) $\log_{10}=8$. We confirm the existence of a prominent peak when using all galaxies (i.e. no cut) right at the scale of the radius of the ring found by HINORA. However, when restricting the data set to brighter and brighter galaxies this peak actually diminishes. Note that the peak seen by us for the complete sample of galaxies does not coincide with the one reported by \citet{Tully1978}, which is at ca. 2~Mpc as opposed to ca. 4~Mpc in our case. Further, the peak in the 2PCF of the LVG galaxies cannot be directly related to our satellite ring that persists even up to a $\log_{10}{L_K}$ cut of 8. To better understand the relation between ring-like structures and peaks in the 2PCF we have performed some additional experiments, again not explicitly shown here: we placed perfect rings into otherwise random data, calculating the resulting 2PCF. We find that this leads to peaks at distances comparable to the inter-point separation on the circumference of the ring and not to peaks at the scale of the radius of the ring. The connection between the rings and the peaks therefore remain to be investigated in more detail. However, when actually reversing the K-band luminosity cuts, i.e. using galaxiese with $\log_{10}{L_K}<[6,7,8]$ we \textit{always} observe the peak at ca. 4~Mpc, the stronger the smaller the applied cut. This reassures us that this feature is driven by the low-mass galaxies in the Local Universe. It appears that low-mass galaxies are correlated on a scale of approximately 4~Mpc, whereas the massive galaxies in the Local Universe arrange themselves as the Council of Giants that lies within the Local Sheet.

\end{document}